\documentclass[referee,a4paper,12pt,traditabstract]{swsc} 

\usepackage{graphicx}
\usepackage{txfonts}
\usepackage{subfigure}
\usepackage{epstopdf}
\usepackage[displaymath,mathlines]{lineno}
\usepackage[authoryear,round]{natbib}
\usepackage[backref]{hyperref}
\usepackage{url}
\usepackage[binary-units=true]{siunitx}

\hypersetup{colorlinks=true,citecolor=cyan,urlcolor=cyan,linkcolor=blue}

\begin{document}

   \title{LUCI onboard Lagrange, the Next Generation of EUV Space Weather Monitoring}

   \subtitle{}
   
   \titlerunning{The LUCI Instrument}

   \authorrunning{West et al.}

   \author{
          M.J. West \inst{\ref{i:rob}, \ref{i:swri}}\fnmsep\thanks{Corresponding author: Matthew J. West \email{mwest@boulder.swri.edu}}
          \and
          C. Kintziger \inst{\ref{i:csl}}
          \and
          M. Haberreiter \inst{\ref{i:pmod}}
          \and
          M. Gyo \inst{\ref{i:pmod}}
          \and
          D. Berghmans \inst{\ref{i:rob}}
          \and
          S. Gissot \inst{\ref{i:rob}}
          \and 
          V B\"uchel \inst{\ref{i:pmod}}
          \and
          L. Golub \inst{\ref{i:harvard}}
          \and
          S. Shestov \inst{\ref{i:rob}, \ref{i:lebedev}}
          \and
          J.A. Davies \inst{\ref{i:ral}}
          }

\institute{
         Royal Observatory of Belgium, Ringlaan -3- Av. Circulaire, 1180 Brussels, Belgium\label{i:rob}
         \and
         Southwest Research Institute, 1050 Walnut Street, Suite 300, Boulder, CO 80302, USA\label{i:swri}
         \and
         Centre Spatial de Li\`ege, Universit\'e de Li\`ege, Av. du Pr\'e-Aily B29, 4031 Angleur, Belgium\label{i:csl}
         \and
         Physikalisch-Meteorologisches Observatorium Davos, World Radiation Center, 7260, Davos Dorf, Switzerland\label{i:pmod}
         \and
         Harvard Smithsonian CfA, HEA MS 58 60, Garden Str, Cambridge, MA, USA\label{i:harvard}
         \and
         Lebedev Physical Institute, Leninskii prospekt, 53, 119991 Moscow, Russia\label{i:lebedev}
         \and
         Rutherford Appleton Laboratory, Harwell Campus, Oxfordshire, OX11 0QX, UK\label{i:ral}
}


\abstract{
LUCI (Lagrange eUv Coronal Imager) is a solar imager in the Extreme UltraViolet (EUV) that is being developed as part of the Lagrange mission, a mission designed to be positioned at the L5 Lagrangian point to monitor space weather from its source on the Sun, through the heliosphere, to the Earth. LUCI will use an off-axis two mirror design equipped with an EUV enhanced active pixel sensor. This type of detector has advantages that promise to be very beneficial for monitoring the source of space weather in the EUV. LUCI will also have a novel off-axis \emph{wide} field-of-view, designed to observe the solar disk, the lower corona, and the extended solar atmosphere close to the Sun-Earth line. LUCI will provide solar coronal images at a 2-3 minute cadence in a pass-band centred on \SI{19.5} {\nano\meter}. Observations made through this pass-band allow for the detection and monitoring of semi-static coronal structures such as coronal holes, prominences, and active regions; as well as transient phenomena such as solar flares, limb Coronal Mass Ejections (CMEs), EUV waves, and coronal dimmings. The LUCI data will complement EUV solar observations provided by instruments located along the Sun-Earth line such as PROBA2-SWAP, SUVI-GOES and SDO-AIA, as well as provide unique observations to improve space weather forecasts. Together with a suite of other remote-sensing and in-situ instruments onboard Lagrange, LUCI will provide science quality operational observations for space weather monitoring. 
 } 

\keywords{
Instrumentation: detectors -- Space vehicles: instruments -- Telescopes; Sun: corona; Sun: UV radiation
}

\maketitle

\section{Introduction}

Space weather is manifest in many forms, from the background solar wind and the interplanetary magnetic field carried by the solar wind plasma, to more short-lived energetic events including \emph{Coronal Mass Ejections (CMEs)}, the sudden release of radiation into space by solar \emph{flares}, and \emph{Solar Energetic Particles (SEPs)} generated by CME driven shocks and flares. Each event can affect the near-Earth system and human life in different ways; SEPs can trigger solar particle events throughout the solar system, where-as the magnetic fields and energetic particles in eruptions and the solar wind can induce geomagnetic storms, ionospheric disturbances and scintillation of radio signals (see \citet{Hapgood2017} for a thorough review). Due to its impacts, especially on ever increasing space-based sensitive equipment, space weather monitoring and forecasting have become increasingly important.

The ESA Space Safety Program (S2P) Space WEather (SWE) segment has been set up to support the utilisation and access of space through the provision of timely and accurate information regarding the space environment, particularly regarding hazards to infrastructure in orbit and on the ground. An important part of the S2P space weather network is space-based instrumentation, where, from the vantage point beyond the Earth's protective atmosphere, space-based instruments can monitor the local environment through in-situ monitoring instruments, as well as make remote sensing observations to anticipate the possible future effects of space weather. 

\begin{figure}
\centering
\includegraphics[width=0.80\columnwidth]{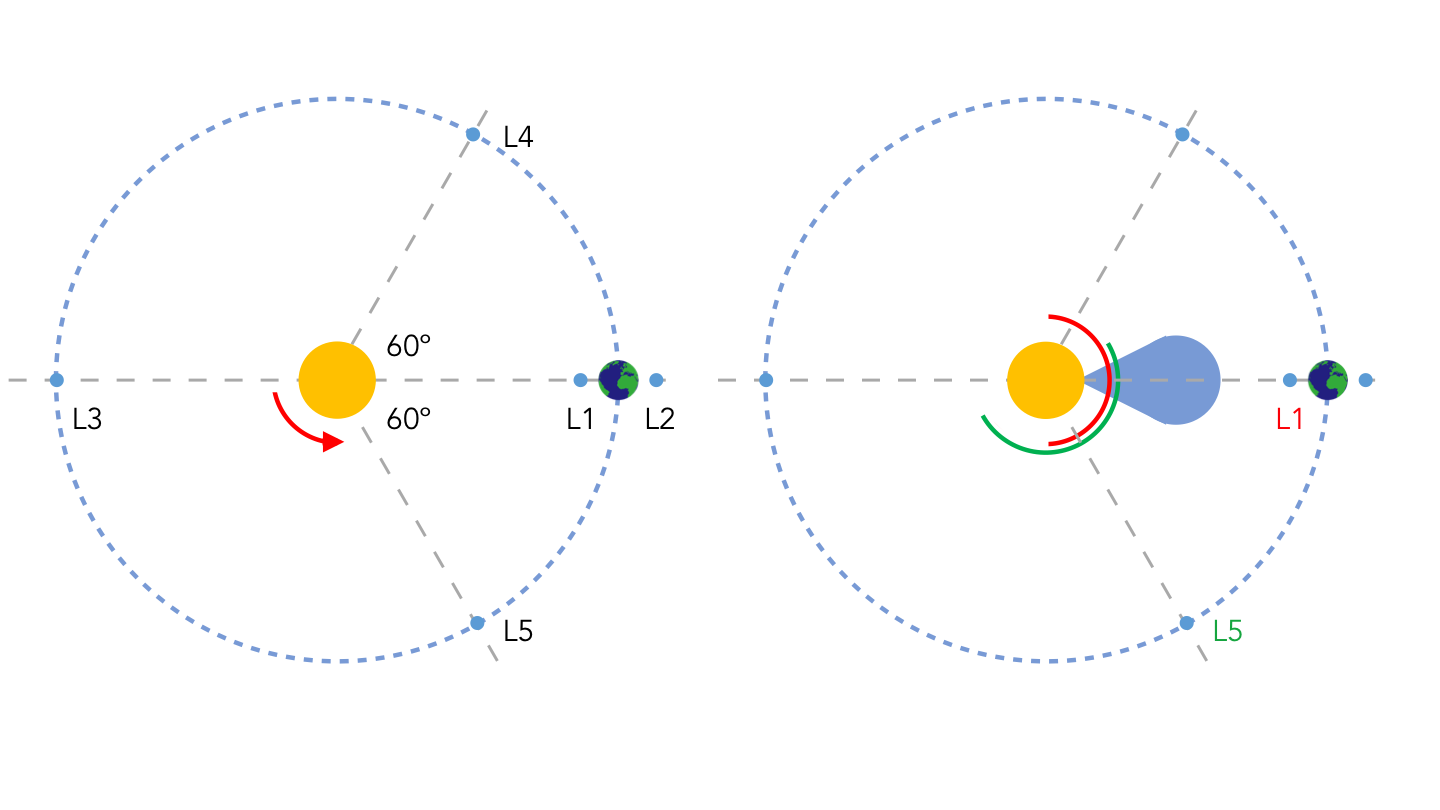}
\caption{\small Left: The position of the 5 Lagrangian points relative to the Sun and Earth system from a solar north perspective, where the red arrow indicates the direction of solar rotation. Right: The different perspectives of the Sun from the L1 and L5 points, represented by red and green lines respectively. The blue ice-cream cone shape represents an Earth directed eruption.} 
\label{fig:LagrangePosition}
\end{figure}

Most space weather monitoring instrumentation is based along the Sun-Earth line, principally in Earth orbit and at the L1 Lagrangian point, which is upstream in the solar wind with respect to the Earth, providing us with information about space weather coming towards the Earth. Another important location identified for space weather monitoring is the L5 Lagrangian point \citep[e.g.][]{Gopalswamy2011}, located approximately 60 degrees behind the Earth in its orbit. The left panel of Figure \ref{fig:LagrangePosition} shows the positions of the 5 Lagrangian points, from a view point above the solar north pole. The Lagrangian points represent positions of relative stability for satellites, where the Sun and Earth's combined gravitational and centrifugal pulls balance. 

An L5 based observatory offers several observational benefits for space weather monitoring. For example, it allows us to monitor eruptions from \emph{the side}, providing more accurate measurements of the direction and speed of phenomena moving towards the Earth, which would otherwise be observed coming directly towards the observer. From the Earth perspective, eruptions emerging from near solar disk centre are of most interest, as it's these events that will eventually reach the Earth. However, they're observed as expanding Halos, making it difficult to measure the speed and size \citep{Gopalswamy2010}, where a wide slow CME may be perceived as a fast narrow CME or vice versa. The blue ice-cream cone shape in the right panel of Figure~\ref{fig:LagrangePosition} represents such an Earth directed eruption. 
 
When monitoring the sources of space weather on the Sun, increasing the size of your view is important, as is increasing the amount of time you can monitor a potential source before it is Earth effective. The red arrow in Figure \ref{fig:LagrangePosition} indicates the direction of solar rotation and highlights the potential for an imager to observe the solar disk before it is Earth orientated. The right panel of \ref{fig:LagrangePosition} indicates the different viewing perspectives of the Sun from an observatory positioned at the L1 and L5 points, represented by red and green lines respectively. An L5 based observatory would extend our view of the Sun by an additional 60 degrees. 

The source of most space weather can be directly attributed to solar phenomena observed in the lower solar atmosphere, on the solar disk and out to several solar radii off the limb. Instruments imaging the Sun through Extreme UltraViolet (EUV) pass-bands provide us with one of the best means of observing this region, allowing us to observe the region through a broad range of temperatures \citep[e.g. EIT on SoHO][]{Delaboudiniere1995}. One example of such an instrument is the Sun Watcher using Active Pixel System detector and Image Processing (SWAP; \citet{Seaton2013,Halain2013}) EUV telescope on PROBA2 \citep[][]{Hochedez2006}, whose operations have been funded through the ESA S2P program for several years. SWAP is a small EUV telescope that images the solar corona, through a pass-band focussed on the Fe IX/X emission lines (corresponding to a \SI{17.4}{\nano\meter} EUV pass-band). EUV observations allow us to image the lower solar atmosphere at temperatures ranging between several thousand and several million K. The fundamental building block of the solar atmosphere is the Sun's magnetic field which permeates the solar atmosphere, and it's at temperatures observed through EUV pass-bands that we can see magnetic structures highlighted by the hot plasma trapped on them. 

\begin{figure}
\centering
\subfigure{\includegraphics[width=0.99\columnwidth]{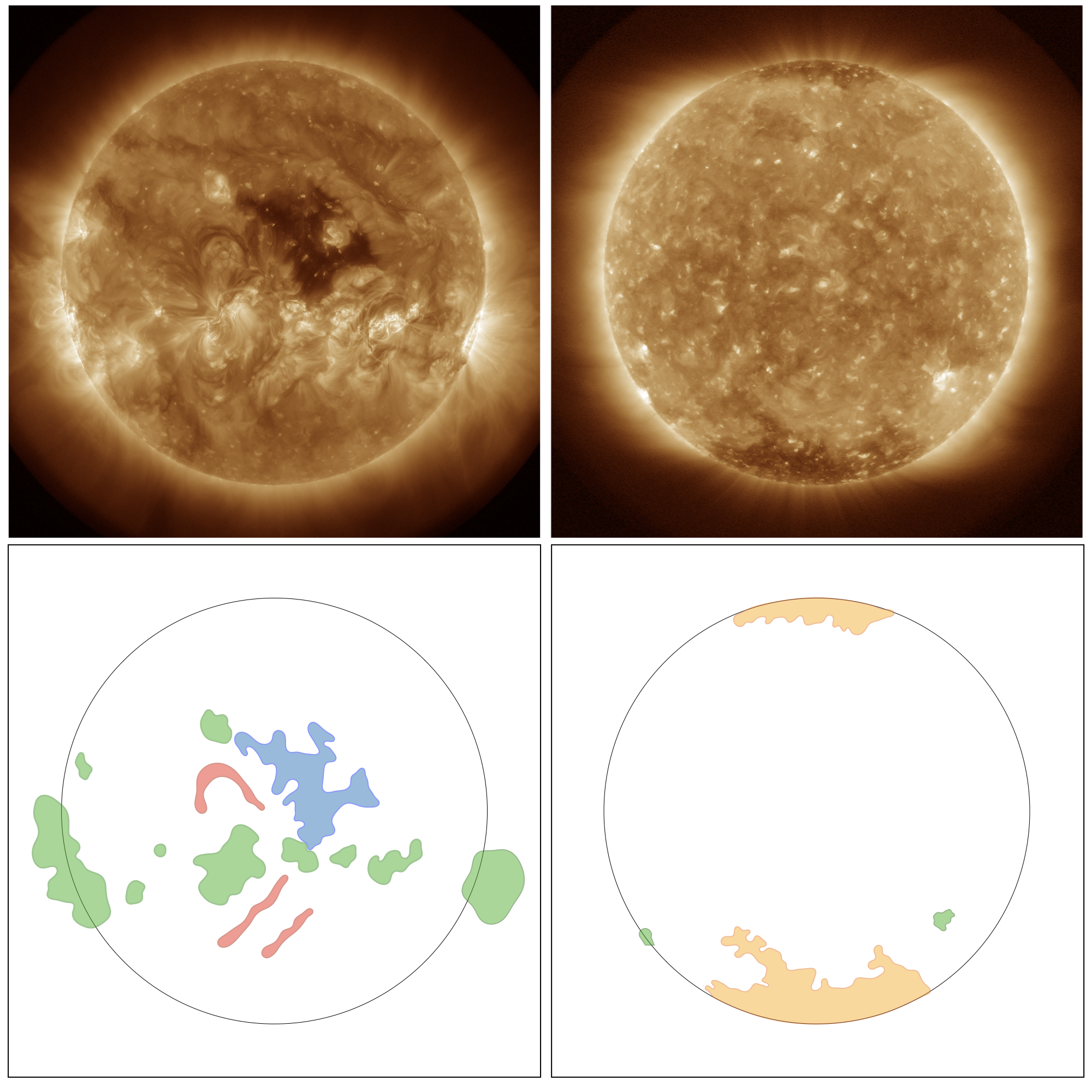}}
\caption{\small Top$:$ Two representative images of the Sun from the AIA instrument on SDO, through the \SI{19.3}{\nano\meter} pass-band, from solar maximum (left; 2014-Jan-01) and solar minimum (right; 2020-Jan-01). Bottom$:$ Highlights several different manually identified regions that can be identified in the corresponding images above, such as active regions (green), equatorial coronal holes (blue), filament channels (red), and polar coronal holes (yellow).} 
\label{fig:aia193}
\end{figure}

Figure~\ref{fig:aia193} shows two images of the solar atmosphere, from 2014-Jan-01 at 00:10 UT (top left) and 2020-Jan-01 at 00:10 UT (top right), taken with the Atmospheric Imaging Assembly \citep[AIA][]{Lemen2012} on the Solar Dynamics Observatory (SDO) in the \SI{19.3}{\nano\meter} pass-band. The AIA \SI{19.3}{\nano\meter} pass-band primarily observes the Sun at temperatures around 1.6~$\times$~10$^{6}$~K. At this temperature we can see the rich menagerie of solar structures, including hot bright \emph{active regions} and the relatively cool \emph{coronal holes}. By comparing the two images in Figure~\ref{fig:aia193} we can also see the stark differences between the structure of the Sun at solar maximum (left) and solar minimum (right), where the solar disk transforms from one being peppered with active regions, indicating regions of intense magnetic activity, to a more placid one dominated by \emph{quiet sun} regions, with two well defined polar coronal holes. The solar activity of interest to the space weather community ranges from long-lived (semi-static) structures, with the aim of trying to anticipate when they may produce activity at the Earth, to more dynamic events. Long-lived structures include coronal holes, active regions and prominences (filaments). More dynamic events, which are often intrinsically related to the long-lived structures, include flares and eruptions. EUV instruments, when combined with coronagraphic white-light observations from instruments such as LASCO \citep{Brueckner1995} onboard the Solar and Heliospheric Observatory (SoHO), allow space weather observers to monitor solar activity from the solar surface out to several solar radii.

The Lagrange mission is being developed under the S2P, SWE segment, to place an operational space weather monitor at the L5 point with the express purpose of monitoring solar activity and space weather along the Sun-Earth line. The Lagrange mission, in its current form, contains four remote-sensing optical instruments and five in-situ instruments to analyse the Sun, inner heliosphere, energetic particle streams, and the ambient solar wind conditions. One of those instruments will be the Lagrange eUv Coronal Imager (LUCI), which is a solar imager in the Extreme UltraViolet (EUV).

Several studies have used observations made by the Sun Earth Connection Coronal and Heliospheric Investigation (SECCHI) package \citep{Howard2008} from NASA's Solar Terrestrial Relations Observatory \citep[STEREO][]{Kaiser2008} mission, when they were positioned close to the L4 and L5 points, to highlight the advantages of monitoring space weather from close to this position. \citet{Bailey2020} showed that predicting the minimum Dst at Earth can be improved with an L5 based observatory, and \citet{Rodriguez2020} showed forecasts of CME arrival times are improved with L5 observational input. \citet{Byrne2010}, using observations from when the STEREO satellites were near quadrature with the Earth ($86.7^{\circ}$ separation), made three-dimensional reconstructions of a CME to determine an accurate arrival time of the eruption at the Lagrangian L1 point, highlighting the efficiency of forecasting with instrumentation off the Sun-Earth line. This work not only utilised white light instrumentation (coronagraphs and Heliospheric Imagers \citep[HI;][]{Eyles2009}), but also EUV observations to help determine the direction of propagation. The work also highlights that significant acceleration occurs lower down in the solar atmosphere, in regions most typically observed with large field-of-view (FOV) EUV imagers.

The LUCI imager is being designed to image the solar atmosphere in the EUV from the L5 Lagrangian point, in an operational capacity with the interests of space weather forecasters in mind. In Section \ref{sec:SpaceWeatherServicesandScience} of this article we review the different phenomena required to be observed by an EUV imager for space weather services and science. In Section \ref{sec:TechnicalInnovations} we present the current design of the instrument, highlighting some of the technical innovations and heritage incorporated into the mission, and section \ref{sec:Discussion} presents a discussion on the rationale behind those decisions, and the logistics of operating an EUV imager in the harsh and remote environment of the L5 point.

\section{Space Weather EUV Observations}
\label{sec:SpaceWeatherServicesandScience}

The Sun is the source of all space weather that could impact the Earth, other planetary bodies and human infrastructure. It therefore needs careful observation to help understand upcoming risks and to forecast the impact. Although the origins of space weather are intrinsically linked to the magnetic fields formed deep in the Sun's unobservable convection zone and tachocline, it's the manifestation in the solar atmosphere that needs to be monitored, in particular in the lower and middle corona. It's in these regions where the motions from within the Sun, combined with large magnetic Reynolds numbers found in the corona, can cause current sheets to build up, and energy to be liberated through the process known as magnetic reconnection. The energy released, and the reconfiguration of the magnetic fields can result in some of the most spectacular space weather events.

The corona is observed to be highly inhomogeneous, which is directly attributed to the magnetic field. In general the solar atmosphere can be separated into open and closed field regions. The closed field regions are predominantly made up of \emph{quiet sun} regions and \emph{active regions}, which are made up of a series of magnetic flux tubes, or loop structures, piercing the photosphere from below. Open field regions trace out magnetic fields that close deep in the heliosphere creating an open appearance on the Sun. 

\begin{figure}
\centering
\subfigure{\includegraphics[width=0.32\columnwidth]{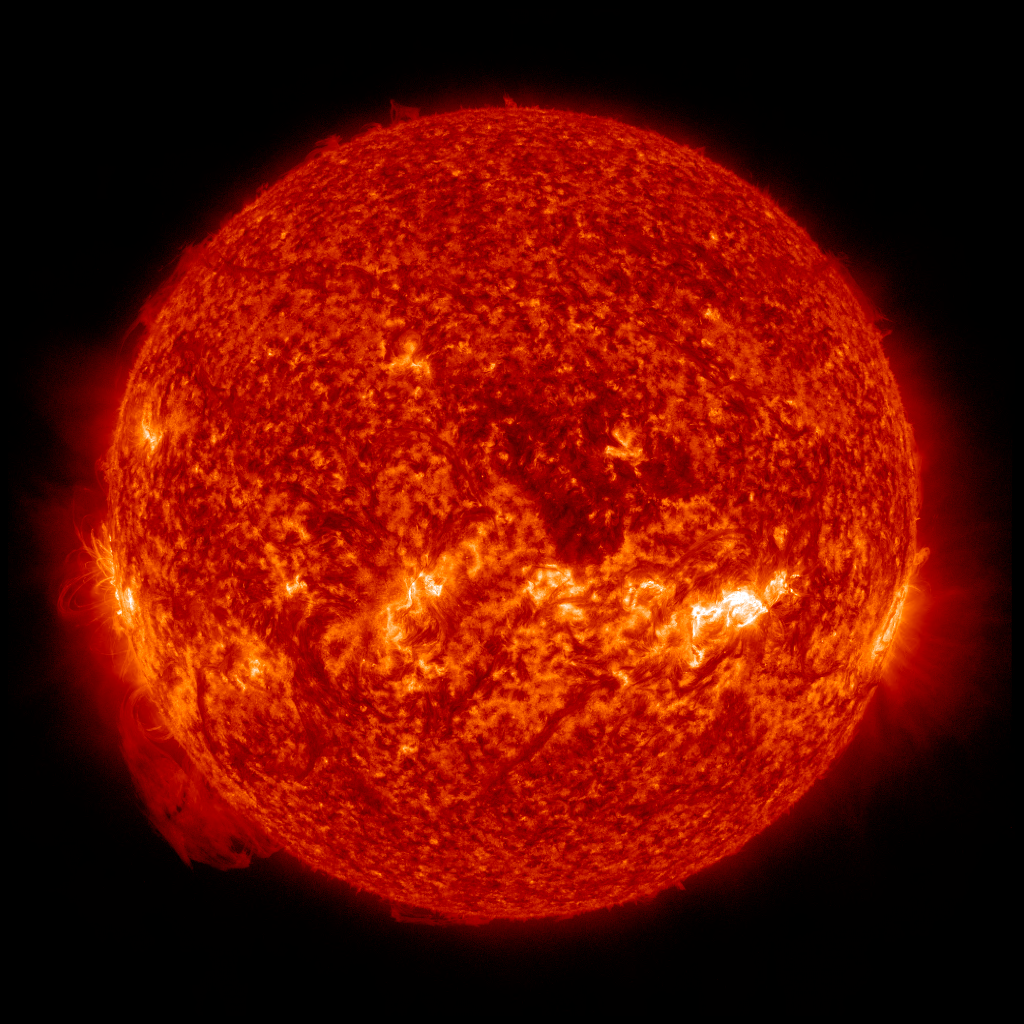}}\subfigure{\includegraphics[width=0.32\columnwidth]{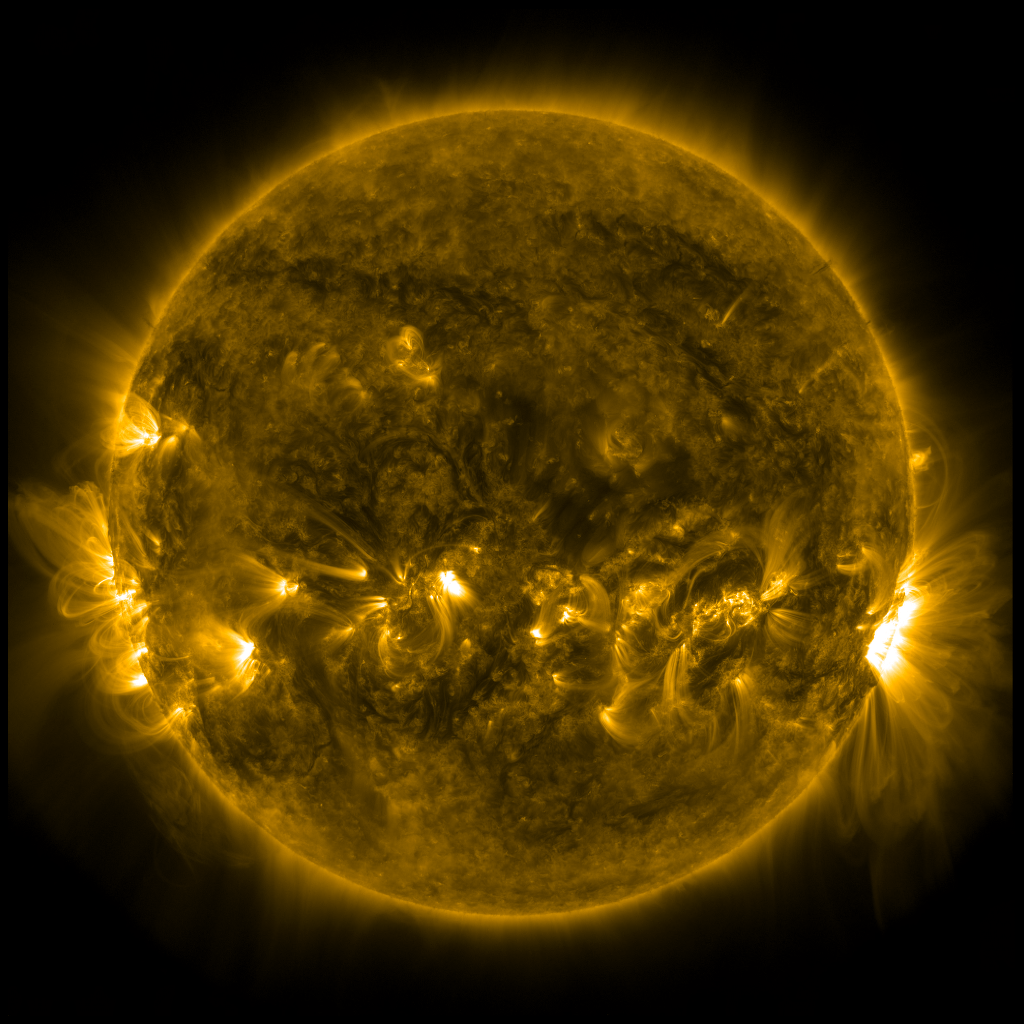}}\subfigure{\includegraphics[width=0.32\columnwidth]{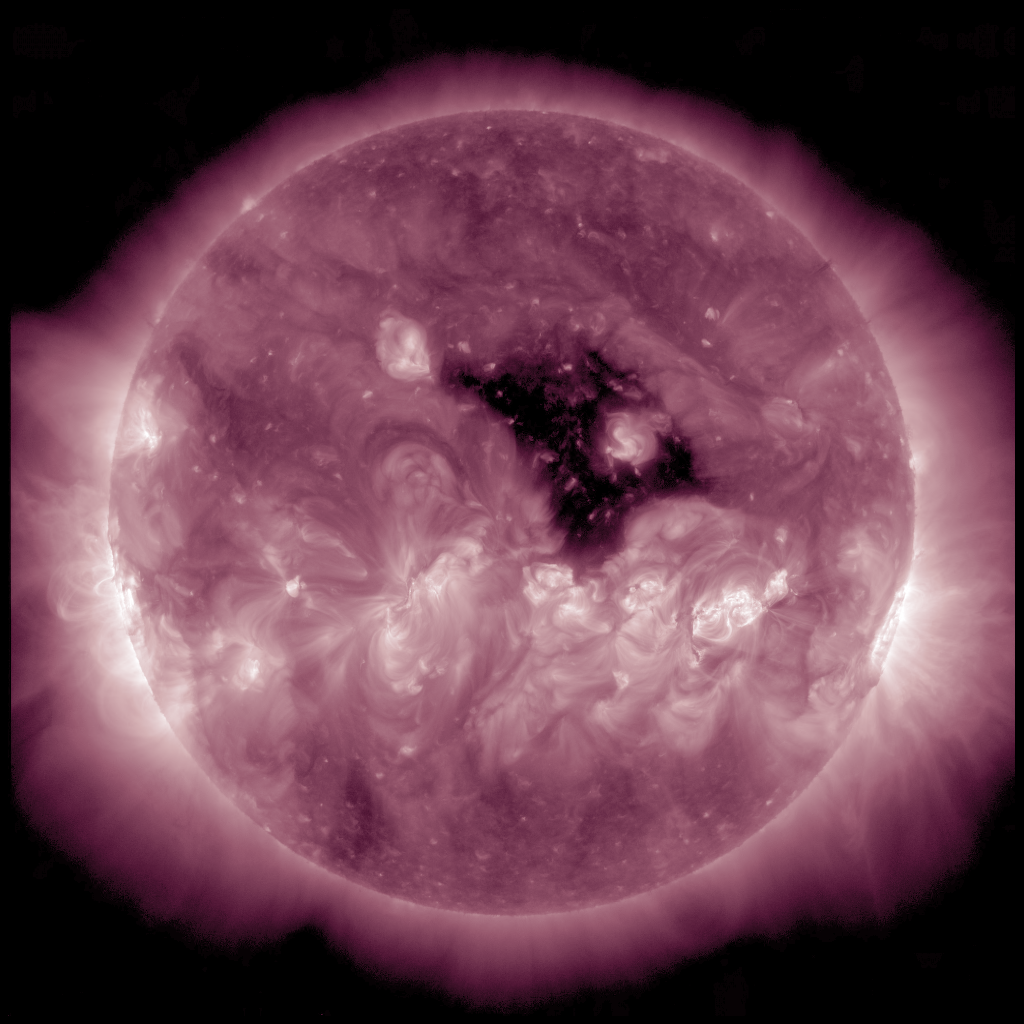}}
\subfigure{\includegraphics[width=0.32\columnwidth]{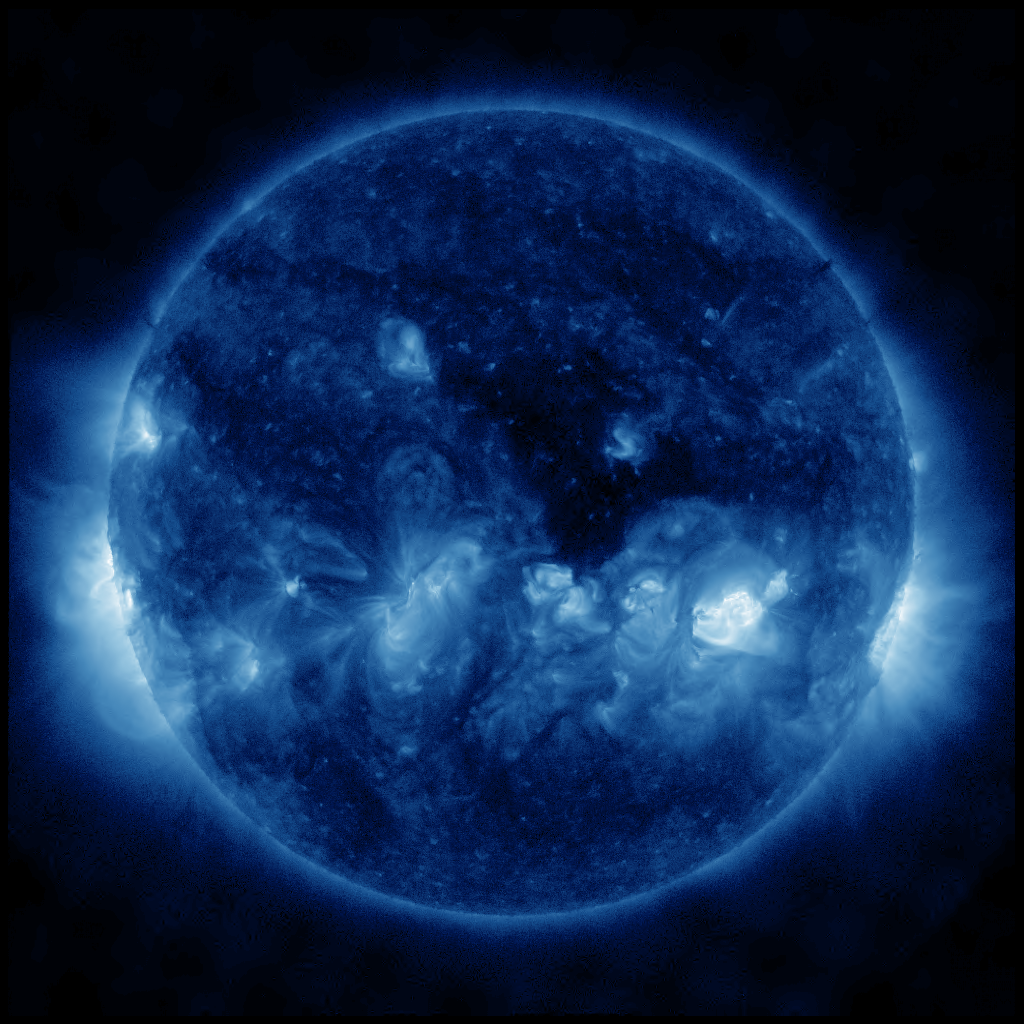}}\subfigure{\includegraphics[width=0.32\columnwidth]{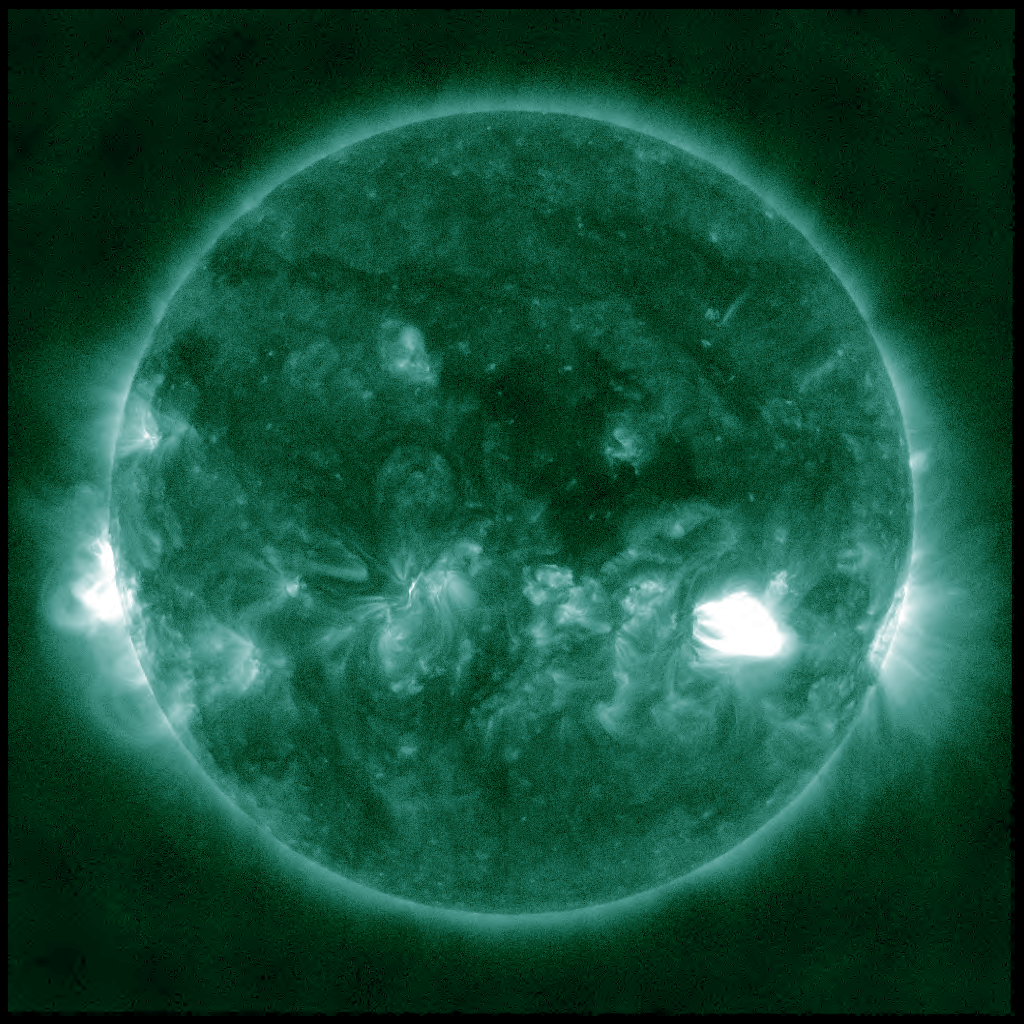}}\subfigure{\includegraphics[width=0.32\columnwidth]{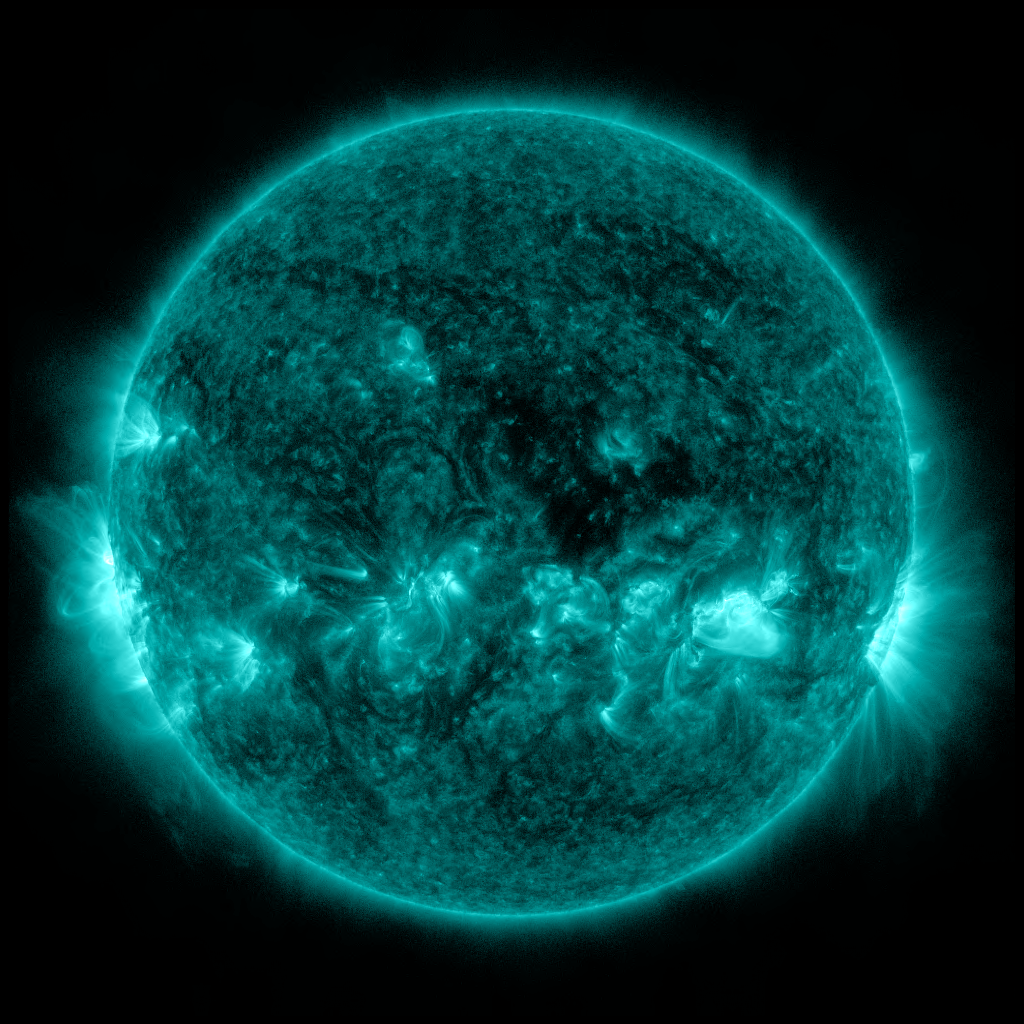}}
\caption{\small The Sun observed in six of the AIA imager's pass-bands on 2014 Jan 01. The top panels (from left to right) correspond to the pass-bands: \SI{30.4}{\nano\meter}, \SI{17.1}{\nano\meter}, \SI{21.1}{\nano\meter}; The bottom row corresponds to: \SI{33.5}{\nano\meter}, \SI{9.4}{\nano\meter}, \SI{13.1}{\nano\meter}. These can be compared to the left panel of Figure \ref{fig:aia193}, which shows the \SI{19.3}{\nano\meter} pass-band from AIA at a similar same time.} 
\label{fig:AIApass-bands}
\end{figure}

The solar atmosphere is best observed in the EUV portion of the electromagnetic spectrum, where hot plasmas ranging in temperature from several thousand to several million Kelvin (K) highlight magnetic structures permeating the region. Due to the range in temperatures and densities of differing structures in the solar atmosphere, observations are filtered into pass-bands, characterized by a limited number of spectral lines. The AIA instrument hosts the largest selection of pass-bands on a single instrument, covering temperatures from chromospheric to coronal regions. Figure \ref{fig:AIApass-bands} shows the Sun in six of AIA's pass-bands, where the top row shows the \SI{30.4}{\nano\meter} (pass-band peak temperature: $5\times10^4$~K; primary pass-band ions: He~II), \SI{17.1}{\nano\meter} ($6.3\times10^5$~K; Fe~IX), and \SI{21.1}{\nano\meter} ($2\times10^6$~K; Fe~XIV) pass-bands, and the bottom row shows the \SI{33.5}{\nano\meter} ($2.5\times10^6$~K; Fe~XVI), \SI{9.4}{\nano\meter} ($6.3\times10^6$~K; Fe~XVIII), \SI{13.1}{\nano\meter} ($4\times10^5~\&~1.0\times10^7$~K; Fe~VIII, XXI) pass-bands. These images can be directly compared with the left panel of Figure \ref{fig:aia193}, which shows an image from the \SI{19.3}{\nano\meter} ($1.6\times10^6~\&~1.6\times10^7$~K; Fe~XII, XXIV) pass-band acquired at a similar time. To say individual pass-bands only observe specific features is disingenuous as many of the coronal structures are multi-thermal and some pass-bands observe broad temperature ranges, through multiple emission lines. However, each of the pass-bands was originally chosen to highlight different structures and regions of the solar atmosphere, such as: the \SI{30.4} pass-band observes the chromosphere and transition region, \SI{17.1} the quiet corona and upper transition region, \SI{19.3} the corona and hot flare plasma, \SI{21.1} and \SI{33.5} active regions, \SI{9.4} the flaring corona, and \SI{13.1} the transition region and flaring corona (see Table 1 in \citet{Lemen2012}).

Teams monitoring space weather are primarily interested in understanding and forecasting when events will happen, and if the events will impact the Earth directly or indirectly. The monitoring can be broadly grouped into two categories, the monitoring of semi-static structures and of dynamic structures. Semi-static structures are long-lived, persisting from days to multiple solar rotations, and include: coronal holes, active regions, and prominences (filaments). Whereas dynamic structures are more short lived, lasting for minutes to hours, and include: flares, CMEs and their related phenomena.

In EUV observations, coronal holes appear as regions of lower emission in contrast to the surrounding corona, due to the lower temperature and density of the plasma observed there. At photospheric and chromospheric temperatures coronal holes are almost indistinguishable from the surrounding corona, however they can begin to be distinguished when the observing pass-band temperature exceeds about $10^{5}$~K. Coronal holes come in two varieties: polar and transient. Polar coronal holes are more persistent, emerging in the polar regions around solar minimum and can persist for many years. In contrast, equatorial (transient) coronal holes \citep{Rust83,Kahler01} can occur anywhere on the Sun, but are more commonly located around the equatorial regions. They're generally more short lived but can still persist for several solar rotations. In the right panel of Figure~\ref{fig:aia193}, two polar coronal holes can be seen capping the top and bottom of the Sun. In the left panel, a large equatorial coronal hole can be seen near disk centre. They are classed as regions of \emph{open} magnetic field, and allow plasma to flow away from the Sun relatively unimpeded, creating High Speed Streams (HSSs) in the solar wind. These HSSs can interact with the relatively slow ambient solar wind, creating compression regions, known as Co-rotating Interaction Regions (CIRs). See \citet{Vrsnak2007a} and \citet{Vrsnak2007b} for a thorough review on coronal holes and HSSs. With an Earth orientation, HSSs and their associated CIRs can enhance geomagnetic activity, the severity of which is dependent on the speed of the stream and the direction of the underlying interplanetary magnetic field. As a consequence the emergence, growth, and development of coronal holes are a main source of interest for teams monitoring space weather, and EUV observations of coronal holes are used to help forecast the arrival of HSSs and CIRs. 

Solar prominences, labelled filaments when seen on the solar disk, are arcade-like-structures seen in absorption at coronal temperatures, as they are made of cool  ($7\times10^3 \le T \le 2\times10^4$~K) dense material, probably of chromospheric origin. They can be seen as thin dark braided lines in several pass-bands with a cooler component, and are best seen at chromospheric temperatures, such as those seen in the \SI{30.4}{\nano\meter} pass-band of AIA (see the top left panel of Figure~\ref{fig:AIApass-bands}). Prominences can be found anywhere on the Sun, but they're often found adjacent to polar coronal holes and active regions. The number of prominences changes with the solar cycle, with more being seen near solar maximum. Prominences can remain stable for several solar rotations, before becoming destabilised and possibly erupting \citep[][]{vanDrielGesztelyi2009} leading to some of the most spectacular eruptions observed. The activation is not fully understood, making forecasts of their eruption challenging. There's some evidence that increased motion inside the foot points can be a precursor to an eruption \citep{Ofman1998} and therefore careful monitoring is required. Monitoring the formation of a prominence and its life cycle is another key EUV observable. See \citet{Parenti2014} for a thorough review of solar prominences.

Active regions are regions of intense magnetic activity; the concentration of magnetic flux emergence, cancellation, and magnetic reconnection, in these regions releases significant amounts of energy, directly and indirectly heating the regions to temperatures in excess of $10^7$~K. The resulting high-density multi-thermal plasma causes them to emit over a broad spectrum, the emission coming from a myriad of closed magnetic loop structures that make up the active region. They are therefore seen in all EUV pass-bands as regions of bright loops (see Figures \ref{fig:aia193} and \ref{fig:AIApass-bands}). The constant reorganisation of the magnetic topology not only leads to a localised release of energy, but it can also produce more dramatic events that are of interest to space weather forecasters, including flares and eruptions. As a consequence active regions are monitored extensively for evidence of upcoming activity. Precursors to activity have been studied extensively; \citet{Zhukov2005} showed that regions exhibiting excessive shear and twist are more eruption productive, whereas \citet{Lee2012} found that active region size is a good indicator of potential flare strength, with larger flares occurring in larger active regions. The decay phase in certain regions can also lead to increased flaring activity \citep[e.g.][]{Su2009}. As a consequence, monitoring the emergence, development (e.g. shape) and subsequent decay of an active region, is paramount. See \citet{Toriumi2019} for a review of active regions and their related phenomena.

Solar flares are one of the most prominent short-lived events monitored by the space weather community, as they produce intense radiation and can produce streams of high energy particles (solar proton events), which can be harmful to spacecraft and astronauts. They effect all layers of the solar atmosphere and are observed at wavelengths ranging from radio waves to gamma rays, and are therefore seen in all EUV pass-bands. Flares can occur anywhere on the Sun, but are generally associated with regions of complex magnetic topology, such as active regions, and it's in these regions where the largest flares are observed. The frequency varies with the solar cycle, from several per day at solar maximum to less than one a week at solar minimum. Flares occur when local magnetic fields reconnect, releasing stored magnetic energy and accelerating charged particles. These particles interact with the plasma medium heating it to tens of millions of K, which in turn can lead to processes such as chromospheric evaporation which can heat coronal loops \citep{Cargill1995}. The evolution of how a flare occurs has been largely laid out in a standard model, which is credited to \citet{Carmichael1964}, \citet{Sturrock1966}, \citet{Hirayama1974} and \citet{Kopp1976}, called the CSHKP model. The model has been built on and improved by numerous authors, developing into the \emph{Flux Cancellation} or the \emph{Catastrophe model} (see e.g. \citet{Lin2000} and \citet{Lin2004}). See \citet{Benz2017} for a thorough review of solar flare observations. Determining the occurrence of a flare is not well understood, however there is some correlation with the magnetic complexity of the source region. As the radiation from flares can reach Earth within minutes, and the exact cause is unknown, forecasts are generally probabilistic, and depend on evidence of previous flaring activity.

Solar flares are sometimes accompanied by CMEs, which are eruptions of magnetized plasma into interplanetary space, and are observed as some of the most dramatic solar phenomena. It is now generally believed that flares and CMEs are two manifestations of a single magnetically-driven mechanism \citep{Webb2012}. Alongside flares, CMEs are of major concern to teams monitoring space weather as they can trigger severe geomagnetic storms. They're generally produced from erupting prominences and from magnetic reconnection in active regions. Although CMEs have been studied for decades, the exact mechanism responsible for their initiation is still under debate, and different eruptions may be produced by different mechanisms. Popular models include the Breakout model \citep[e.g.][]{Antiochos1999}, internal tether-cutting reconnection models \citep[e.g.][]{Moore2001}, general flux cancellation models \citep[e.g.][]{Martens2001}, and ideal MHD kink/torus instability models \citep[e.g.][]{Kliem2006}. See \citet{Webb2012} for a thorough discussion. 

\begin{figure}
\centering
\includegraphics[width=0.85\columnwidth]{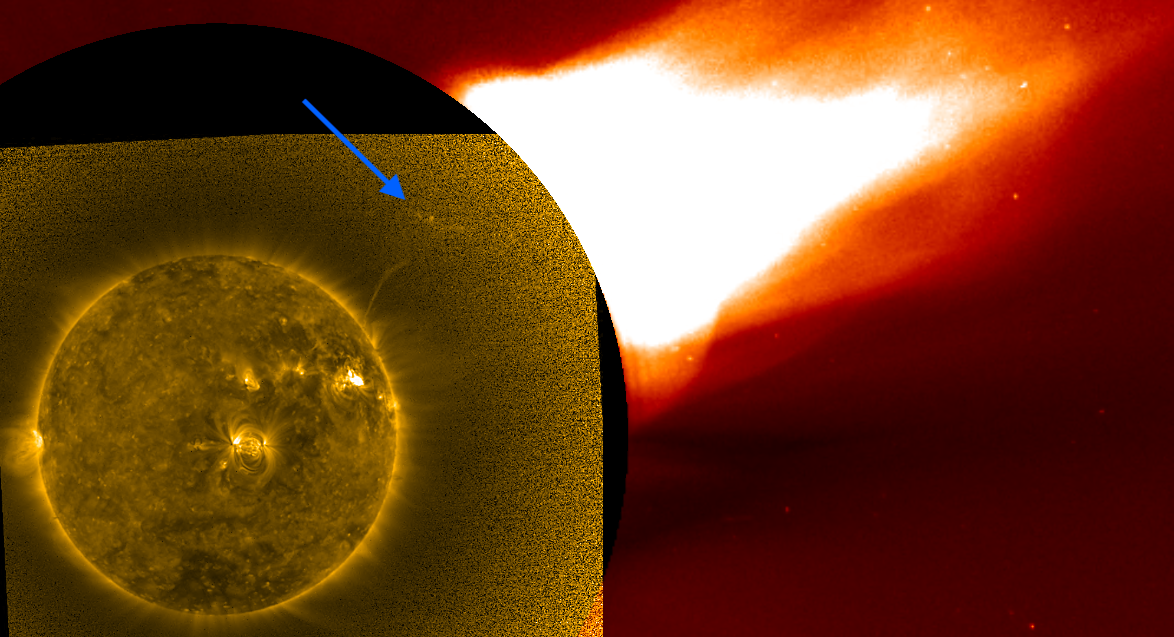}
\caption{\small A LASCO white-light coronagraph image from 2017 Apr 01 at 22:12 UT, with an off-pointed SWAP EUV (\SI{17.4}{\nano\meter} pass-band) image overlaid in the centre (from 22:07 UT). The signature of an eruption can be seen extending out to beyond 1 solar radii above the solar limb in EUV observations (indicated by a blue arrow), and the corresponding CME signature can be seen in the LASCO white-light observation above it.} 
\label{fig:SWAPoffpoint}
\end{figure}

Like flares, predicting eruptions is difficult, although there appears to be some correlation with the amount of magnetic shear and twist in the source region, and also with the age of prominences \citep{Zhukov2005}. However, unlike flares, CMEs can take several days to reach Earth, and careful measurements of their kinematics close to the Sun can allow forecasters to predict if they will hit the Earth, and, if so, when. Measurements are often made in white-light observations (e.g. coronagraphs), but the source region and direction is often determined from EUV observations. Eruptions are often described as having three phases: (1) initiation, (2) main acceleration, and (3) propagation \citep[e.g.][]{Zhang2006}. The first two phases are governed by the Lorentz force whilst, in the later phase, the drag force becomes dominant \citep[e.g.][]{Cargill1996}. The main acceleration phase occurs lower down in the solar atmosphere, below the FOV of most coronagraphs, and therefore is often not characterised well \citep{Byrne2014}. However, in recent years there has been a push to measure this phase with large FOV EUV imagers, like SWAP. \citet{OHara2019} used unique SWAP off-point observations to close the gap with the inner edge of the LASCO coronagraph FOV and was able to track a CME from the solar surface out into the coronagraph FOV. An image from the SWAP campaign, from 2017 April 1, can be seen in Figure \ref{fig:SWAPoffpoint}, overlaid on a LASCO white-light image. In general, EUV images from the L1 orientation are not used by the forecasting community to measure CME kinematics, as Earth effective eruptions are generally located close to solar disk centre and are difficult to discern from the background solar disk due to the optically thin nature of the region. However, observations from an L5 perspective would provide forecasters and modellers with valuable acceleration information and an early warning of potential CME arrival times at Earth. As with flare predictions, CME predictions are mainly probabilistic. Therefore, space weather forecasts will often depend on the monitoring of previous eruptive activity from a region.

EUV imagery is often used to observe several other lower coronal phenomena such as \emph{EUV waves}, \emph{coronal dimmings}, bright \emph{shock fronts}, and \emph{post-eruption arcades}. These signatures are often the byproduct of an eruption \citep[e.g.][]{Podladchikova2009} and can therefore be used to identify potential source regions in EUV imagery. Coronal dimmings and EUV waves are best observed in pass-bands observing around 1-2 million K, such as the \SI{21.1}{\nano\meter} and \SI{19.3}{\nano\meter} pass-bands \citep{Kraaikamp2015}.

Coronal dimmings are transient coronal holes appearing on timescales of minutes to hours, and they are most commonly associated with CMEs. When occurring close to the limb they map well with the foot points of a CME \citep{Webb2000}. As they're seen in multiple EUV pass-bands they're often interpreted as density depletions in the EUV corona due to the evacuation of mass associated with the CME \citep{Zhukov2004}. The presence of a coronal dimming is a good indicator for determining the source of an eruption in EUV imagery.

Although the exact nature of EUV waves is unknown, they are often observed propagating away from the foot point of an eruption with speeds ranging between $50-1500$~km~s$^{-1}$ \citep{Thompson2009}. The bright fronts propagate freely through the quiet Sun but can be seen to reflect off strong magnetic structures such as active regions and coronal holes \citep{Ofman2002}. Theories describing the nature of EUV waves include: blast waves, compression driven fast magnetosonic waves (or shock waves) \citep{west2011}, and EUV waves being the ground tracks of the CME expansion itself \citep{Liu2014}. Waves offer a good indication of the direction of an associated eruption.

Post-eruptive loop arcades are often seen in the lower solar atmosphere in the wake of an eruption, they're described in the aforementioned CSHKP standard model as being generated from magnetic fields ejected from the reconnecting plasma in the wake of a CME, where particles are accelerated into the lower solar atmosphere, heating the plasma which fills newly created and surrounding magnetic loop arcades \citep{Hudson2001}. The post eruptive loop systems can be extensive and persist for days \citep{West2015} and are observed as multi-thermal structures in all EUV pass-bands, providing another indicator for the source of an eruption.

The visibility of each of the above mentioned phenomena in EUV observations will vary depending on the chosen pass-band and the phenomena's temperature and density, especially in contrast to the surrounding atmosphere. An EUV observatory positioned at the L5 Lagrangian point would serve two purposes: to track semi-static structures, and to offer important information on the position of coronal holes and the location of eruptive phenomena that might be Earth effective. The LUCI instrument has been designed to monitor each of the above mentioned phenomena. 

\section{Instrument Overview}
\label{sec:TechnicalInnovations}

The LUCI instrument is being designed to observe the full EUV solar disk and a portion of the off-limb solar atmosphere, through a single channel, from the L5 point. Such a design is challenging due to the harsh radiation environment, mass, power, and the telemetry budget imposed. As an operational mission, the instrument is being designed to last for 6-7 years producing an image every 2-3 minutes, with low risk of operational or mechanical failure. The instrument design takes significant heritage from both the SWAP instrument on PROBA2 and the EUI \citep{Rochus2019} on Solar Orbiter. Following both designs has the advantage of using tried and tested technology and reducing the risk of failure. At the time of writing the SWAP instrument has been operating as a science and operational instrument from low Earth orbit for over 10 years with minimal spectral degradation experienced, and less than 9\% of its pixels have failed. Instrument design, calibration and observing campaigns can heavily rely on the experience gained from these missions. In this section we outline the current design for the LUCI instrument. 

\begin{figure}
\centering
\includegraphics[width=0.95\columnwidth]{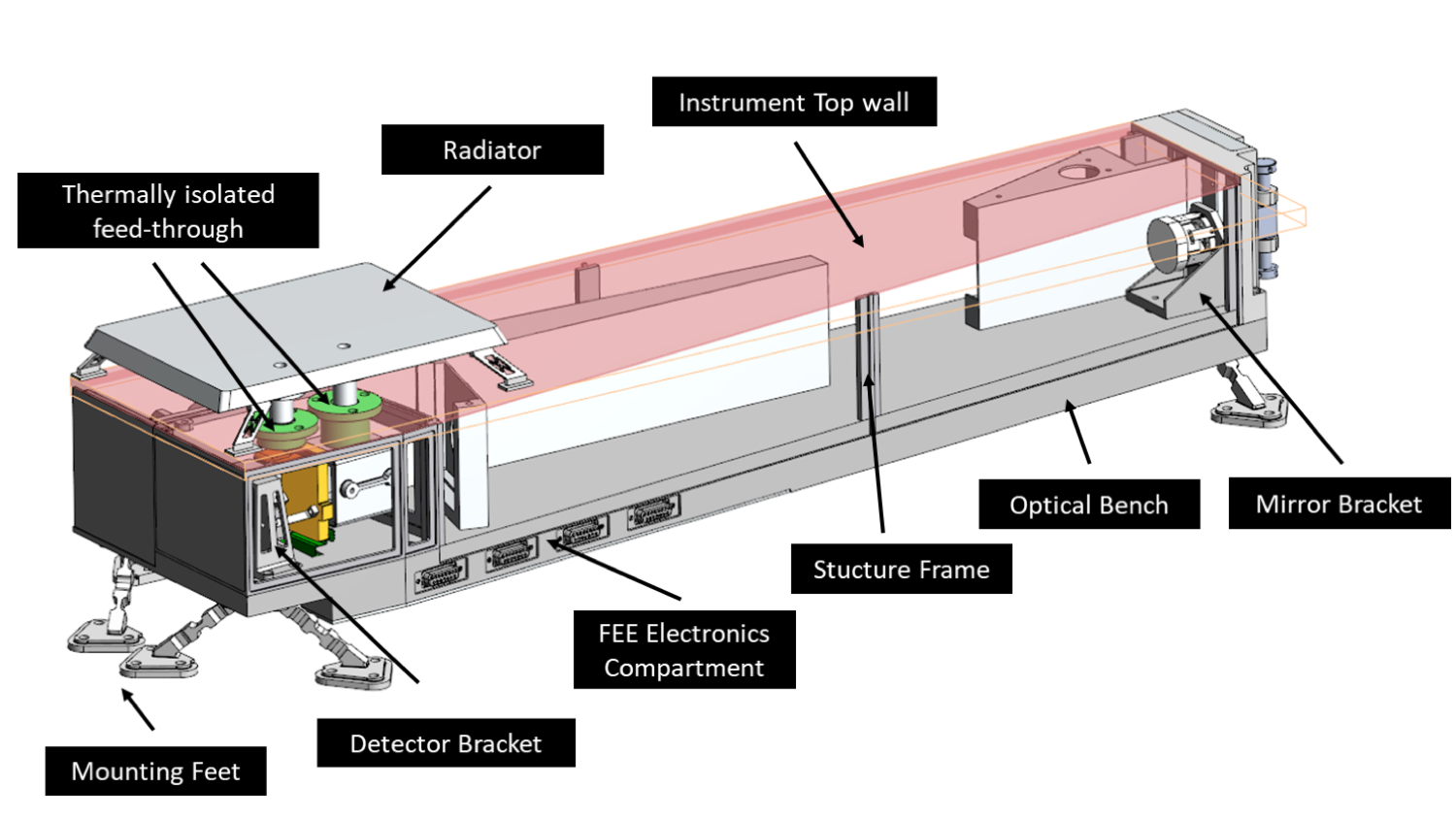}
\caption{\small LUCI optical unit design overview, showing the over-arching structure, including the relative position of each of the components.} 
\label{fig:DesignOverview}
\end{figure}

\begin{figure}
\centering
\includegraphics[width=0.95\columnwidth]{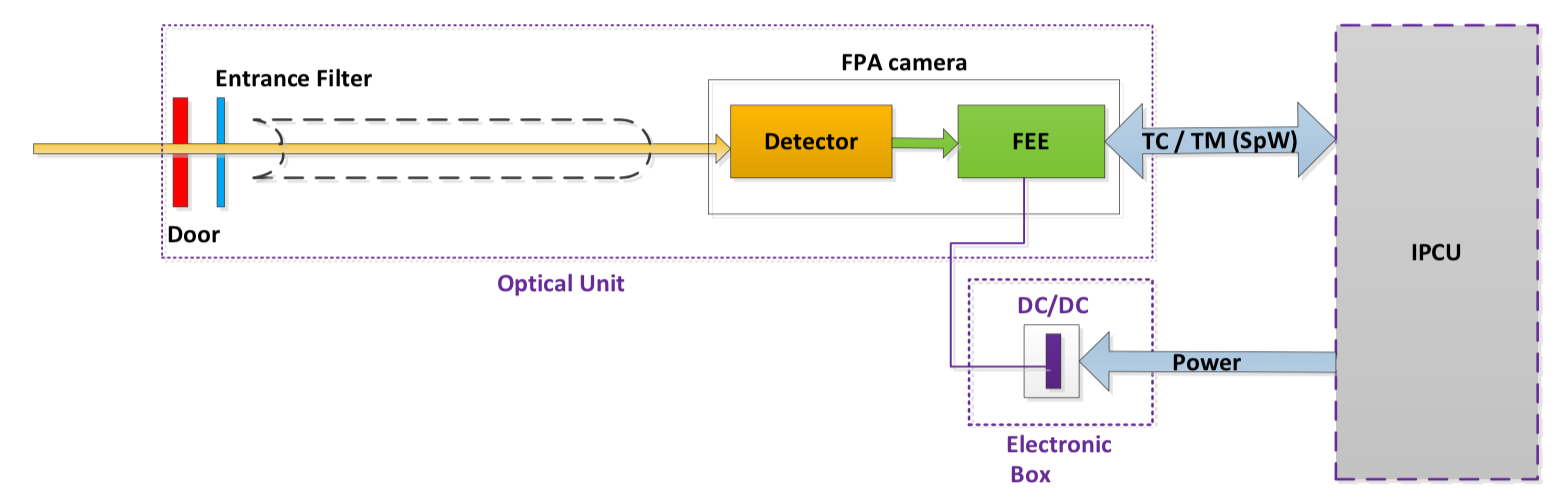}
\caption{\small The LUCI functional architecture, highlighting the connection between the optical unit, electronic box and the Instrument Processing and Control Unit (IPCU).}
\label{fig:FunctionalArchitecture}
\end{figure}

The LUCI telescope is being designed with a two-mirror off-axis scheme. An overview can be seen in Figure \ref{fig:DesignOverview}. The optical design uses heritage from the SWAP instrument, which was designed to minimize the telescope size and to allow for a simple and efficient internal baffling system, where the absence of a central obstruction allows for a smaller aperture of 33~mm. See \citet{Defise2007} for further details. LUCI will be primarily comprised of an optical unit and mounted electronic box. The optical unit implements a single channel optical system for imaging the Sun onto a detector (including a primary parabolic off-axis mirror and a secondary flat mirror). The optical unit contains visible and infrared light blocking filters, optical baffles, and a detector assembly with dedicated camera for image acquisition, called the Focal Plane Assembly (FPA). The functional architecture can be seen in Figure~\ref{fig:FunctionalArchitecture}, and the optical prescription of LUCI is given in Table~\ref{ref:LUCI_design_parameters}.

The instrument will observe through a single pass-band, centred on \SI{19.5}{\nano\meter}, imaging the Sun on to the FPA. The spectral selection will be achieved with EUV reflective multilayer coatings deposited on the mirrors, together with aluminium foil filters that reject visible light and infrared radiation. A Mo/Si coating, similar to that used on SWAP, will probably be implemented due its observed low degradation with time. The overall stack is specifically designed to provide reflectivity in the EUV range and to achieve the spectral selection in a narrow pass-band (1.5 nm at full width half maximum). The accuracy of the central wavelength will be within $\pm$~\SI{0.2}{\nano\meter}. 

The FPA will include a dual-gain Active Pixel Sensor (APS) CMOS (Complementary Metal-Oxide-Semiconductor)  detector, a cold cup for contamination control, decontamination heaters, the Front End Electronics (FEE), and a conductive link to an external radiator for thermal management. The CMOS pixel size will be \SI{10}{\micro\metre} and consequently the focal length will be $\approx$~\SI{1289}{\milli\metre}. The compact design of LUCI is estimated to have a length, width and height of \SI{960}{\milli\metre}, \SI{310}{\milli\metre}, and \SI{195}{\milli\metre} respectively, weighing under  $<$~15~kg, and will have a total power budget of 10~W. LUCI will not have a dedicated computer, and will communicate with the central onboard computer, called the Instrument Processing and Control Unit (IPCU) via the FEE through a Space Wire (SpW) interface. A separate redundant SpW interface will also be implemented. The electronic box will contain the instrument power converter for the FEE as well as some housekeeping electronics.

LUCI is being designed in a way to minimise risk of failure, and as such, several precautions have been taken, these include the use of qualified electronic parts with a high radiation tolerance or immunity, the optical unit and electronic box will be shielded by $>1$~mm of aluminium, and a single one-shot entrance door for protection from contamination during delivery and launch. Furthermore, the survival heaters will be operated by the IPCU in case the instrument is not powered. One survival heater will support the electronic box, and a second heater will support the thermal interface on the FPA, to ensure the temperature stays higher than a predefined threshold required for the sensor to function. The heater will be used periodically for decontamination, to outgas any condensation layers that build up on the cold sensor. This is essential, as molecular deposits can absorb EUV radiation and dramatically decrease photon throughput. Finally, the FEE is mounted to the bottom of the optical bench, and will be electro-magnetically shielded to prevent interference with other systems.

\begin{table*}
\begin{tabular}{lll}
\hline\hline
Focal length                                  & $\approx$~\SI{1289}{\milli\metre} \\
Entrance pupil                                & $\approx$~\SI{33}{\milli\metre}   \\
Detector (full)                                  & \num{3072 x 3072}, \SI{10}{\micro\metre} pixels \\
Detector (nominal window)             & \num{2300 x 1600}, \SI{10}{\micro\metre} pixels \\
Field of view (nominal window)      & 61.3~$\times$~42.7 arcmin \\
Plate scale                                      & \SI{1.6}{arcsec/pixel} \\
pass-band                                       & \SI{19.5}{\milli\metre}  \\
Main Ions (peak ion temperature)  & Fe~XII (1.6~MK), Fe~XXIV (16~MK), Ca~XVII (6.3MK) \\
Volume                                          &  \SI{960}{\milli\metre}, \SI{310}{\milli\metre}, and \SI{195}{\milli\metre} (length, width, height) \\
Weight                                          & $<$~15~kg\\
\hline
\end{tabular}
\caption{Nominal characteristics of the LUCI channel and instrument parameters. }
\label{ref:LUCI_design_parameters}
\end{table*}

\subsection{The Optical Unit and Radiator}
\label{sec:TheOpticalUnit}

The number of reflections in the LUCI optical design is limited to two, to optimize the optical throughput. The mirrors will be manufactured from Zerodur, or a fused silica equivalent, with super-polished optical surfaces, and the mirror brackets will also be primarily constructed of aluminium. All optical elements will be mounted onto an aluminium optical bench, which will support the unit. Aluminium is being used as the primary material to help electromagnetically isolate the telescope from the satellite. The optical bench will be mounted onto the spacecraft via six isostatic titanium feet on three mounts, preventing optical misalignment in case of spacecraft deformation. These can be seen extending from the undercarriage of the optical bench in Figure \ref{fig:DesignOverview}. The optical unit and electronic box will be equipped with an isolated feed that will allow for the measurement of resistance between the unit and the structure. Bonding to the structure will be performed via studs and all metallic parts will be electrically conductive, and surface treated where possible to improve electrical conductivity. A secondary ground will be connected to the structure through the electronic box.

\begin{figure}
\centering
\includegraphics[width=0.95\columnwidth]{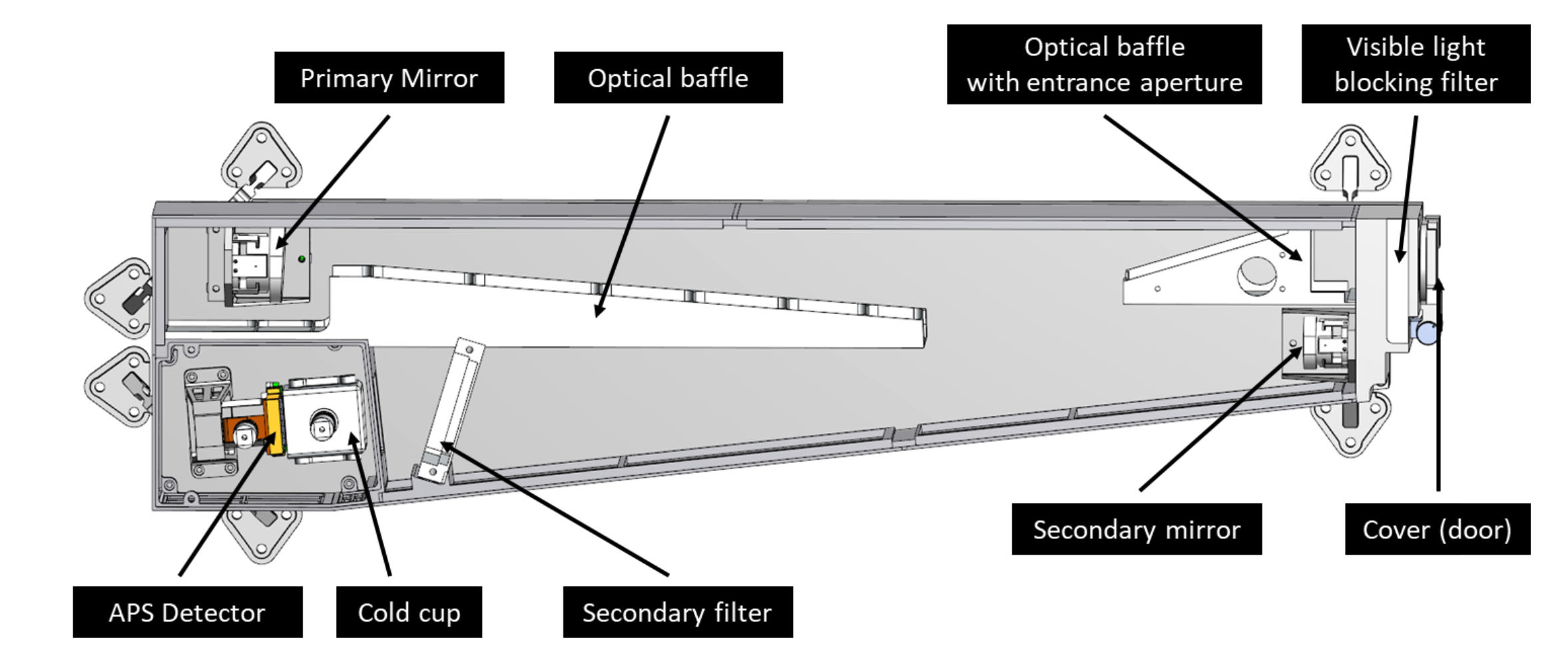}
\caption{\small LUCI instrument Top view.} 
\label{fig:InstrumentTopView}
\end{figure}

In order to prevent LUCI from being illuminated by visible and infrared light, rejection filters will be introduced within the optical path of the telescope, see Figure \ref{fig:InstrumentTopView}. Filters similar to those implemented on SWAP will be used, allowing 80\% EUV transmission. These filters will need to be supported by a mesh with a thin polyimide membrane, which offers a good transmission rate, with the drawback of the entrance filter creating a diffraction pattern on the detector and the focal plane filter creating a shadowing effect. Both the effects of the diffraction pattern and shadowing can be reduced with image-processing \citep{Auchere2011}. An achromatic transmission of 82\%, and an overall maximum transmittance of 14.2\% on the detector is anticipated. The photon flux falling onto a given pixel is calculated by integrating the pixel spectral flux with respect to the wavelength over the selected pass-band. With the proposed LUCI optical design, photon rates of 14513, 812, and 49 ph~s$^{-1}$~pixel$^{-1}$ are estimated from active regions, the quiet sun, and coronal holes respectively. Through provisional estimates of the detector efficiency (9.7~e$^{-}$ / UV photon) and electronic gain (5~e$^{-}$ / DN), we calculate corresponding signal-to-noise ratios to be 534, 126, and 29 respectively, where the noise is estimated from a combination of photon shot noise, noise contribution from the analog-to-digital converter (including quantisation noise), readout noise of the detector and noise of the dark current.

A stray-light prevention system of two planar optical baffles will be implemented in order to avoid any direct view between the detector and the entrance pupil, and vanes will be added where necessary to prevent stray-light from indirect sources.

An aluminium radiator will be attached to the optical unit via four titanium elements, which are optimized for low thermal conductivity. The radiator, which is located above the FPA (see Figure \ref{fig:DesignOverview}), is supported by titanium blades that allow for differential dilatation. The thermal feed-throughs isolate against the instrument top wall, will allow axial movement to a certain degree, and include seals to minimize the contact area whilst protecting against contamination. 

Both the primary and secondary mirrors have a common mount design of three titanium blades fixed on a bracket which is glued on the mirrors. This interface allows shimming for mirror alignment. The front filter is mounted close to the door mechanism to reduce the acoustic vibration pressure on the filter. Venting holes with labyrinths placed close to the front filter allow depressurisation and outgassing and reduce the risk of differential pressure on both sides of the filter, either during depressurisation or due to acoustic vibrations.

Several calibrated diodes will be placed within the optical unit and used periodically to help characterise the detector. Periodic calibration campaigns are envisaged whereby the satellite is off-pointed, to remove substantial EUV signal and stray-light from the detector. The diodes will then be used to characterise the detector for bias (offset), dark-current corrections, and to help identify underperforming pixels. Maps (look up tables) of the corrections will be constructed on the ground and uploaded to the satellite for onboard correction.

\subsection{The Focal Plane Assembly and Electronic Box}
\label{sec:FPA}

\begin{figure}
\centering
\includegraphics[width=0.95\columnwidth]{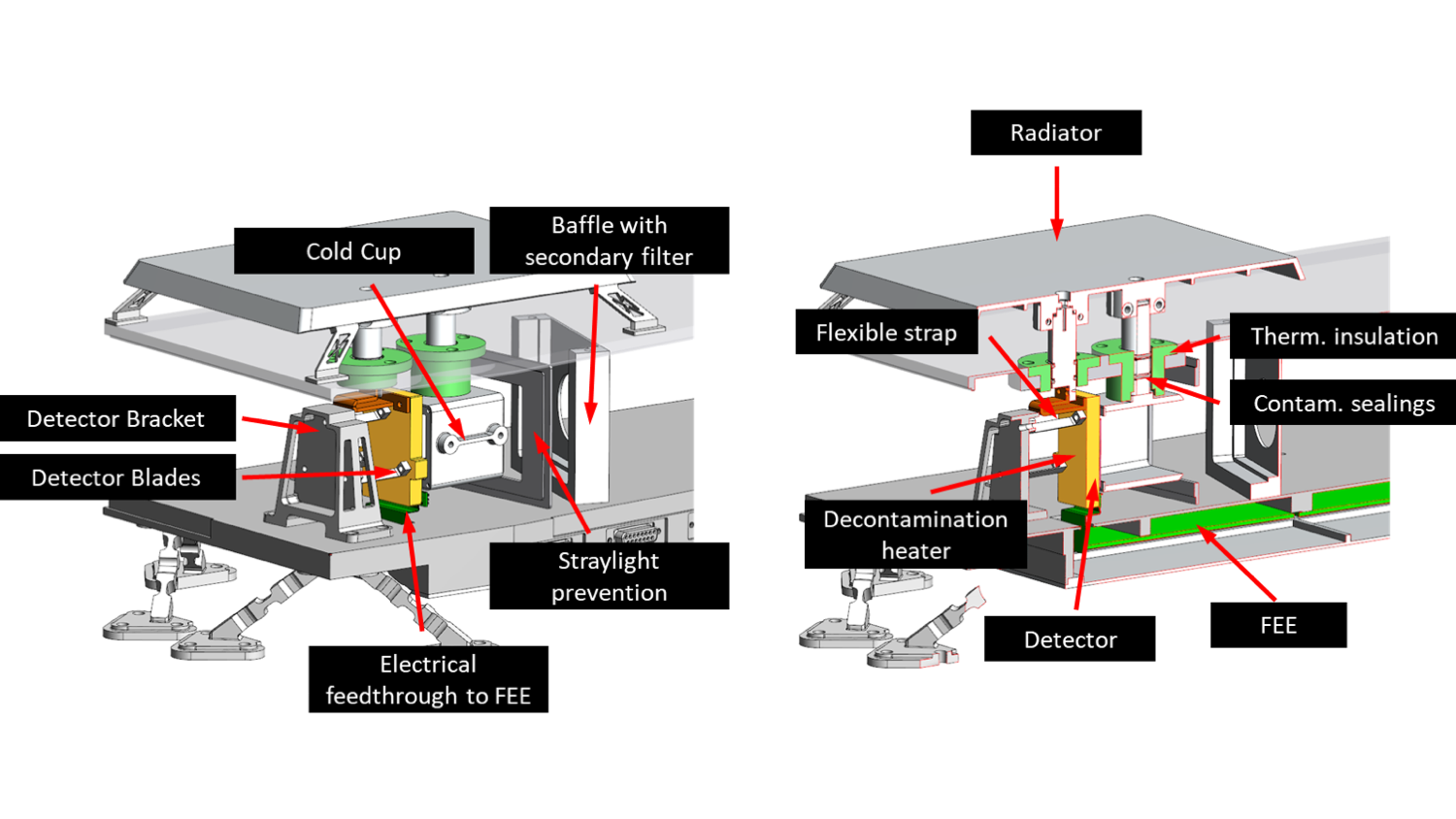}
\caption{\small LUCI Focal Plane Assembly.} 
\label{fig:LUCIFPA}
\end{figure}

The FPA is comprised of an APS-CMOS detector, a decontamination heater, a cold cup, the FEE, and a thermal link to the external radiator for cooling the detector down. Figure \ref{fig:LUCIFPA} shows a section of the FPA. The cold cup is designed to trap contamination and is situated in front of the detector, where a high thermal conductive connection ensures the cold cup is the coldest point within the FPA. The FEE is situated on the underside of the optical bench. This location prevents possible contamination of the detector by the electronic board and assures short distances to the detector readout electronics. Decontamination heaters are mounted on the backside of the detector, and will be used after launch to support out-gassing and bake-out campaigns, when required. These heaters will be controlled by the IPCU with the aid of local thermistors located on the detector and cold-cup.

The EUV photons will be collected on an APS-CMOS detector, in a backside thinned configuration. The detector has been procured from the same detector set as those used on EUI (see Section 5.1 of \citet{Rochus2019} for full details). The full detector accommodates 3072~$\times$~3072~pixels of \SI{10}{\micro\metre} each. For each pixel, the signal is proportional to the solar flux corresponding to the small viewing angle of the pixel. The electrical signals are converted to digital numbers in the FEE. To increase the dynamic range of each image, a dual-gain pixel design has been developed where each pixel outputs both a high and a low gain signal, allowing the acquisition of images with an improved dynamic range. The high gain channel has low read noise and a low saturation level, which is desirable for observing the faint off-limb corona. The low gain channel has a high saturation level but larger read-noise for observing the bright solar disk.

Similar to the SWAP instrument, the detector is expected to be covered by a scintillator coating (approximately \SI{10}{\micro\metre} thick). The phosphorous coating, known as “P43” (Gd$_{2}$O$_{2}$S which gets activated by Tb), absorbs EUV radiation and re-emits it as visible light (at \SI{545}{\nano\meter}) to which the APS-CMOS is sensitive; see \citet{Seaton2013} for further details. The coating has been selected due to its observed low degradation.

The LUCI electronics mainly consist of the FEE and an electronic-box with power unit. The FEE is required to read out the sensor, and is connected by a SpW interface to the IPCU. The FEE is divided into two parts, an analogue and a digital part. The analogue part reads out the image sensor, which is converted into a digital signal, to be transferred to the IPCU. The electronic box contains a low power voltage interface (DC/DC), line filter, the power converter for the FEE, some housekeeping electronics, and a power interface to the IPCU. The unit is supported by survival heaters.

Thanks to the APS-CMOS design, a rolling shutter system is implemented where transfer can start after imaging of the first row has finished. Rows of pixels are read off the APS-CMOS detector, and the FEE will select the high or low gain channel on a pixel-by-pixel basis dependent on a predefined signal threshold. The threshold will be changeable via individual tele-command, or through a pre-defined observation commanding scheme. The signal from each pixel will be 14-bits, with an additional 15th-bit attached to indicate the gain channel from which the pixel was selected. Due to telemetry restrictions only a window of 2300~$\times$~1600~pixels of the full 3072~$\times$~3072~pixels will be read out, producing a wide-FOV image of 61.3~$\times$~42.7 arcmin (see section \ref{sec:LUCIdata} for further details). Assuming a net transfer rate of 70~Mbit via the SpW interface, an image will take approximately 0.79~s (0.92~s with overhead) to transfer to the IPCU, where it will undergo pre-processing and compression.


\subsection{Onboard Software}
\label{sec:OnboardSoftware}

\begin{figure}
\centering
\includegraphics[width=0.99\columnwidth]{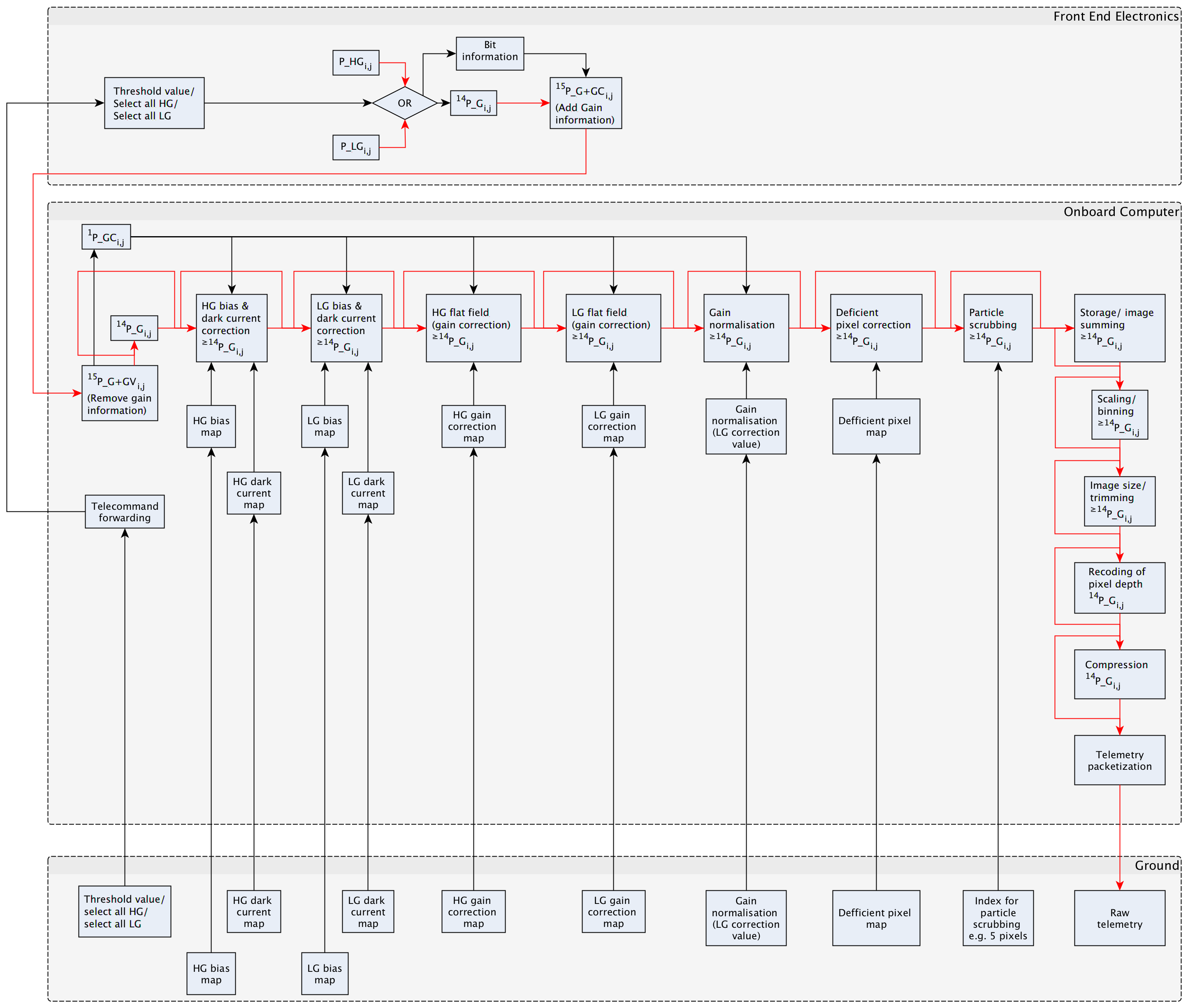}
\caption{\small The onboard processing flow required to process images before transmission to the ground, where P represents a pixel within an image. P$_ {i,j}$ represents data acquired from a pixel on the detector located at position $i,j$. The red arrow indicates the operational flow. The black arrows indicate additional inputs to each section of the on-board software. HG indicates a pixel from the high gain channel. LG indicates a pixel from the low gain channel. G indicates a pixel in a combined HG, LG image. GC indicates the gain channel, e.g. 1 for high gain and 0 for low gain. } 
\label{fig:onboardSoftware}
\end{figure}

Image processing will be performed on a separate computer, such as the IPCU, where the image of 15~bit pixels will be received from the FEE. The first step in the image processing chain will be to remove and store the gain indication bit in a separate map, which will be required at a later stage for gain correction. The remaining steps in the processing chain are concerned with preparing the acquired image for transmission to the ground. Each of these steps will be optional and configurable from the ground. They include: a bias subtraction, an offset/dark-signal (non-uniformity) correction, a deficient pixel removal scheme, energetic particle scrubbing, a small-scale flat field, image size scaling, summing, compression, and a recoding of the pixel depth. Each of these steps can be seen in the processing diagram in Figure~\ref{fig:onboardSoftware}, where the red arrows indicate the overarching operational flow. It is foreseen that each step in the processing flow should be optional via tele-command and can therefore be skipped if needed.

With an APS-CMOS detector, each pixel is treated individually, with its own offset and dark-signal. Subtraction of a dark map of coefficients, prepared and uploaded from the ground, will be performed to correct for any such offset. Each channel of the detector will require its own map. The maps will be updated throughout the mission and characterised through calibration campaigns. 

A gain (small scale flat-field) correction may be required to normalise the signal from each pixel; the correction will serve two purposes. First, a pixel-by-pixel gain correction for either gain channel may be required to correct for non-uniformities between individual pixels. Second, if the image contains a combination of pixels from the high and low gain channels, the low gain signal should be corrected with a pre-defined factor to normalise the signal with those acquired from the high gain pixels. The gain correction is treated as a multiplicative factor. Two gain correction maps (one for each gain) will be required, which will, like the dark map, be updated throughout the mission, and characterised through calibration campaigns.

Out of the millions of pixels in an imaging sensor, there can be quite a number with a deviating behaviour, that can range from absolutely dead (zero output signal) to continual oversensitivity (signal always saturated). If the deviation is strong enough and persistent in every image, the pixel value is in fact meaningless and will hamper the compression. The deficient pixel removal scheme assumes a 1~bit pixel map onboard that has the same physical dimensions of the sensor. This pixel map is configurable from the ground, and will be updated throughout the mission. When an image is passed through the deficient pixel removal scheme, each pixel in the sensor for which the bit is set in the deficient pixel map, will be replaced by the average or median of its neighbours.

Besides pixels that have a value deviating persistently in every image, it is also possible to have enhanced brightness in single pixels (occasionally streaks) created during the image acquisition by the FEE, due to the impact of charged particles from cosmic sources, or solar particle events. Such outliers will appear in single images but they can equally disturb the compression. A particle scrubbing technique will primarily be used to remove artefacts. Outliers are identified by comparing them with their neighbours spatially (and optionally temporally) through a scheme such as the Median Absolute Deviation (MAD) algorithm, a method to estimate the local standard deviation in the presence of outliers. Any flagged pixel is then replaced by a median of the surrounding pixels. If temporal comparisons are to be applied, then multiple images will need to be stored in the onboard memory.

It is foreseen that the nominal image acquisition will be a single image with no summing required. However, an option to sum images will be reserved in the case that EUV signals are weak in the far-field of an image, and need to be enhanced. The number of images will be adjustable via the observation commanding scheme. If summing is implemented, multiple images will need to be stored in the onboard memory. Adaptive schemes, where only portions of the image are summed may also be applied.

Image scaling (binning) will not form part of the nominal image acquisition pipeline, but is reserved for potential high cadence campaigns, or the production of low resolution images, whereby the acquired image is scaled down to reduce telemetry. For example, full detector images (3072~$\times$~3072 pixels) may be read from the FEE and scaled down to half resolution to preserve telemetry. If required, pixels will be binned X~$\times$~X, where X could be a factor of 2 or greater. Assuming the pixels are summed in a 2~$\times$~2 pattern, omitting the under-scan and over-scan pixels, an image comprised of 14-bit signals could be stored as a 16-bit unsigned integer image without information loss.

As described above, LUCI will be capable of acquiring observations across its full 3072~$\times$~3072 pixel detector. In nominal acquisition mode a window of the full detector will be acquired and read out by the FEE. A second image trimming may be applied through the IPCU. Such trimming may come into effect if pointing knowledge acquired during image acquisition shows the image to include unwanted regions. Trimming could therefore be implemented to save telemetry. Although trimming and binning are conceptually separate, they will be performed as part of the same process since the trimming is just an offset in the binning origin and end point.

The final stage of the software processing chain is pixel recoding, in which each pixel is scaled between user-defined low and high values. The photon detection process suffers from an intrinsic uncertainty due to its quantum nature. This can be exploited by applying a non-linear mapping on the recorded pixel values, which leads to a bit-depth reduction while maintaining the signal dynamic range and reducing noise, which should improve the performance of any compression algorithm; see \citet[][]{Nicula2005} for further details. Recoding will be required if large gain corrections are applied. For example, the program could quickly replace (recode) all 16 bit values into 14 bit values.

Before final telemetry packetization, the images will be compressed. Both lossless and lossy compression can be applied, as required by the telemetry budget and the acquisition mode. CCSDS standards will be followed, probably using the CCSDS 123.0-B-2 standard. In the case of lossy compression, different limits may be applied to the quantizer to compress different portions of an acquired image by different amounts (e.g. on-disk to on-limb variations). A configurable pixel-by-pixel mask, which decides which maximum error is applied, may be used. 

\subsection{LUCI Expected Output}
\label{sec:LUCIdata}

\begin{figure}
\centering
\includegraphics[width=0.99\columnwidth]{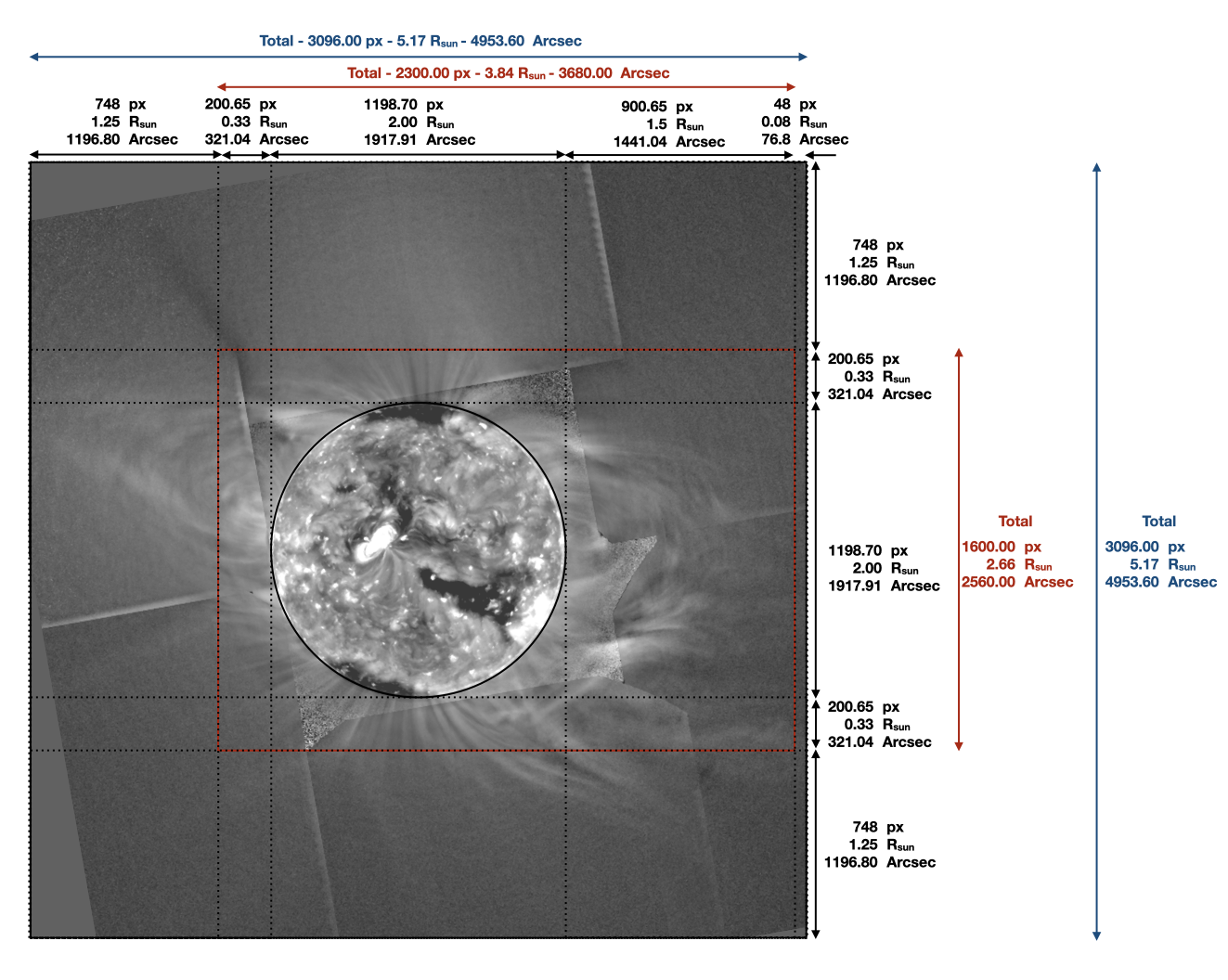}
\caption{\small Represents the FOV of LUCI. The background image is made up of seven SUVI images acquired during a mosaic campaign and stitched together. The image has been scaled and trimmed to the proposed LUCI resolution and FOV. The images were taken through SUVI's \SI{19.5} {\nano\meter} pass-band. The red box indicates the nominal LUCI window, with proposed pixel and length scales included.} 
\label{fig:LUCIFOV}
\end{figure}

The LUCI instrument will provide images of the solar disk and the extended solar atmosphere through a pass-band centred on the \SI{19.5} {\nano\meter} wavelength, corresponding to light produced by highly ionised Fe~XII, with a peak temperature of 1.6~$\times$~10$^{6}$~K. LUCI will be capable of producing 3072~$\times$~3072 images in both high and low gain channels. Nominally, a window of the full detector will be acquired, trimming regions of lower interest, and ultimately reducing the overall telemetry. The image will be composed of a combination of high and low gain pixels. A nominal image will be 2300~$\times$~1600 pixels, with a plate-scale of 1.6~arcsec per pixel, producing a novel wide-FOV image of 61.3~$\times$~42.7 arcmin. Figure \ref{fig:LUCIFOV} shows a SUVI mosaic image, comprised of seven background subtracted  images, taken through a pass-band centred on \SI{19.5} {\nano\meter} on 2018-Feb-12 at around 04:00 UT (see \citet{Tadikonda2019} for further details). The image has been scaled and trimmed to represent the full FOV of the LUCI detector. The red box indicates the nominal LUCI window, with proposed pixel and length scales included. 

\begin{figure}
\centering
\includegraphics[width=0.95\columnwidth]{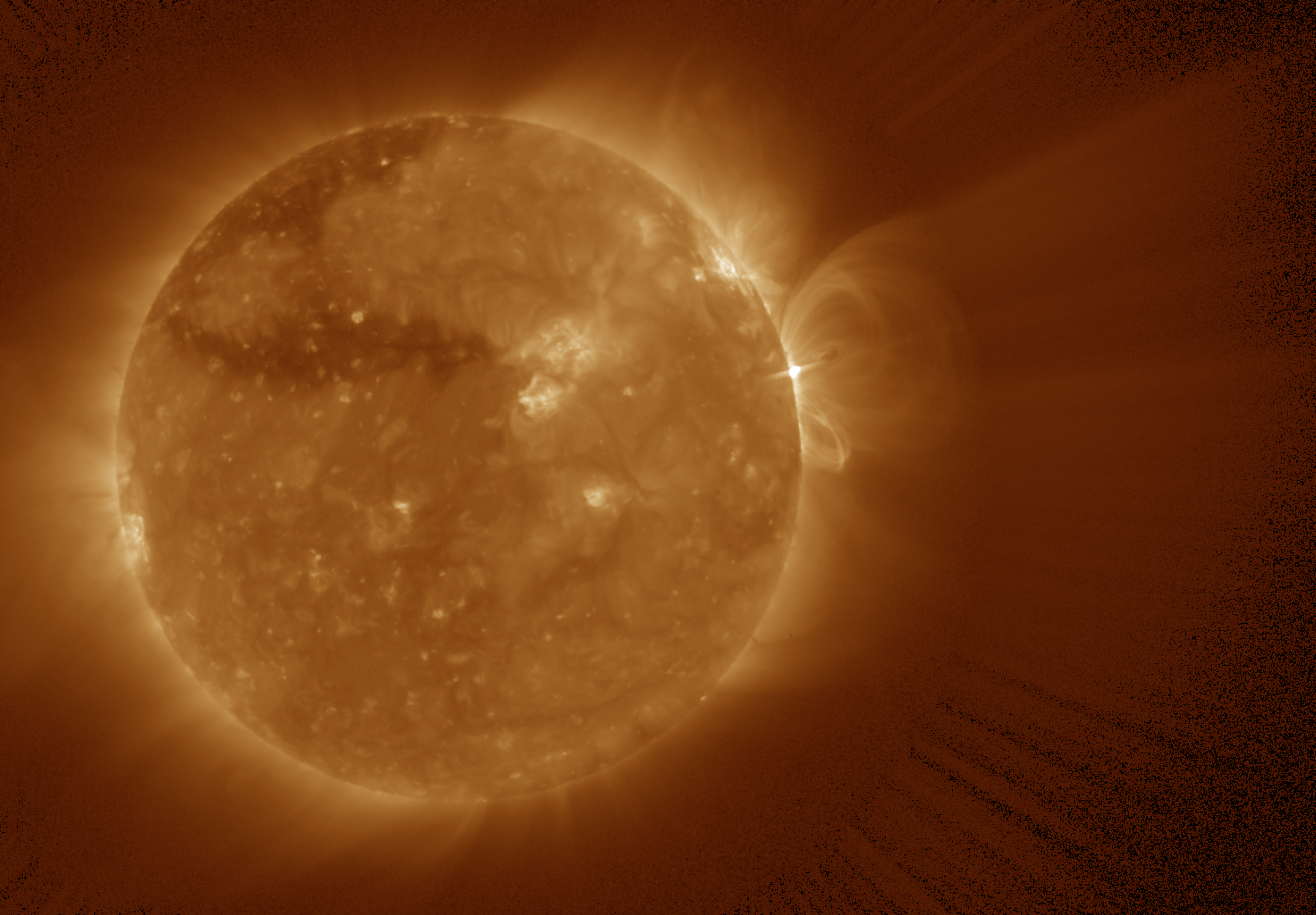}
\caption{\small A representative reconstruction of an anticipated LUCI image, using processed SUVI \SI{19.5} {\nano\meter} pass-band observations.} 
\label{fig:LUCIimage}
\end{figure}

The full LUCI FOV is designed to reach the inner edge of an average coronagraph FOV, filling the observational gap between the coronagraph and the EUV imager scenes. The Sun centre will always be in the middle of the full FOV array. However, the nominal observation window will trim the image to the top, bottom and left of the Sun, restricting the FOV to 5.3~arcmin of solar atmosphere in these directions, whilst the FOV to the right of the Sun (close to the Sun-Earth line from an L5 perspective) will extend to 24~arcmin from the solar limb (1.5 Solar Radii from the solar limb). Figure \ref{fig:LUCIimage} shows an artificial representation of what LUCI can expect to observe. The image is constructed from a SUVI image observed on 2017-Sep-10, through the \SI{19.5} {\nano\meter} pass-band. SUVI's FOV is 53.3~$\times$~53.3 arcmin, therefore, the image was trimmed on the top, bottom and left, whilst the right portion of the image was extrapolated out to the LUCI FOV.  

As described in Section \ref{sec:FPA}, the instrument will implement a high and low gain signal reconstruction. In principle, the high gain channel will be used to image the faint off-limb solar atmosphere, whereas the low gain channel will be used to image the bright solar disk, in regions where the high gain channel will saturate. The low gain signal will then be adjusted with the onboard gain correction (see Section \ref{sec:OnboardSoftware}) by a pre-defined factor, to normalise the signal. The factor will be determined through pre-flight and in-flight calibration campaigns, allowing it to be adjusted throughout the solar cycle to compensate for the changing brightness of the EUV Sun. By adjusting the threshold, it will also allow the instrument to create individual high gain or low gain images.

\section{Discussion}
\label{sec:Discussion}

The Lagrange mission is a planned operational mission to monitor the Sun and solar-wind from the L5 Lagrangian point. LUCI has been designed to be the EUV instrument onboard Lagrange. The logistical challenges of the mission environment have impacted most of the decisions that have gone into the instrument design. Here we discuss the merits and reasoning behind several of those decisions, and how they have shaped the LUCI instrument design.

\subsection{Pass-Band Selection}
\label{sec:pass-bandSelection}

The logistical constraints of sending an instrument to the L5 point led to the decision to use a single channel instrument design. The instrument needs to be of compact design, requiring a low power consumption, and have a low mass. Building upon the heritage of SWAP, an off-axis two mirror design has been decided on, producing an instrument of modest length, width and height (960~mm, 310~mm, and 195~mm respectively), weighing under 15~kg, and with a total power budget of approximately 10~W. It was originally considered that LUCI would have several observational pass-bands, which would have been implemented through multiple instruments, or with a filter wheel design. However, the logistical constraints reduced the design to a single unit, and to reduce additional mechanical risk, a filter wheel has not been included. It should be noted, that at time of writing, EUV imagers: EIT, EUVI on STEREO~A, and AIA, have all been operating successfully without a failure in the filter wheel mechanism, for over 24, 13, and 10 years respectively.

The decision to use a single pass-band is reinforced by the restricted telemetry budget available. The addition of a second or third channel would have led to several compromises in individual pass-bands, such as a reduction in resolution, cadence, or FOV. A lower resolution would have compromised the ability of the detector to monitor smaller scale structures, such as flare location; a reduction in cadence would have reduced the ability to temporally identify the sources of flares and eruptions, as well as inhibited the ability to monitor off-limb eruptions. Finally, a reduction in the FOV would have reduced the ability to monitor the off-limb solar atmosphere, one of the primary observables of the mission.

Figures \ref{fig:aia193} and \ref{fig:AIApass-bands} show images of Sun through the most commonly used pass-bands available to forecasting teams. The pass-band centred on \SI{19.5} {\nano\meter} was selected as it covers some of the brightest spectral lines emitted by iron ions in the solar atmosphere, including Fe~XII (\SI{19.512} {\nano\meter}, with a peak temperature of T$=$1.6~$\times$~10$^{6}$~K). It will also observe highly ionised Fe~XXIV (\SI{19.203} {\nano\meter}; T$=$1.6~$\times$~10$^{7}$~K) and Ca~XVII (\SI{19.285} {\nano\meter}; T$=$6.3~$\times$~10$^{6}$~K). Similar pass-bands have been successfully used on other instruments such as EIT, AIA, and SECCHI-EUVI, and following a poll of several space weather forecasters, a preference for the \SI{19.5} {\nano\meter} pass-band was expressed, as it reveals most relevant phenomena. 

The temperature range observed by the \SI{19.5} {\nano\meter} pass-band is sufficiently broad to capture most space weather related phenomena (see Section \ref{sec:SpaceWeatherServicesandScience}), and is ideal for monitoring both the quiet corona and hot flare plasmas, where a strong signal is generated from the Fe~XXIV and Ca~XVII lines. Images generated from similar pass-bands show good contrast for tracking coronal holes, limb eruptions, EUV waves and coronal dimmings. The only phenomena not easily observed through this pass-band are prominences (filaments) due to their characteristic temperatures ($10^3-10^4$~K; \citet{Parenti2014}), where they are seen as dark threads in absorption. Prominences are better monitored through chromospheric observing pass-bands such as the \SI{30.4} {\nano\meter} channel (see Figure \ref{fig:AIApass-bands}), but 30.4 is not satisfactory for monitoring other space weather related phenomena, as it does not reveal sufficient details of active region dynamics, which are better observed through hotter pass-bands. 

Flares have a broad multi-thermal component and can be readily identified in \SI{19.5} {\nano\meter} observations (as in most pass-bands). However, to identify source regions more precisely, and to gain a better understanding of the magnitude of the flare, hotter pass-bands are better suited, such as the \SI{9.4} {\nano\meter} channel, which was designed to observe the hot corona. However, other phenomena such as EUV waves, dimmings and coronal holes have weaker signals in the hotter pass-bands, producing poor contrast with the background corona. The \SI{13.1} {\nano\meter} channel on AIA is another channel with a hot component. It is a dual peak pass-band observing Fe VIII normally and Fe XXI during energetic events, allowing the observer to monitor cooler and hotter material simultaneously. Unfavourably, EUV waves, and eruptions emitting around a couple of million K, exhibit poor contrast in this passband. Another pass-band of interest is centred around the Fe~XI at \SI{18.8} {\nano\meter} line, which is used on the EUV imaging spectrometer on Hinode \citep{Culhane2007} and SPIRIT \citep{Shestov2008, Shestov2014}, which has strong sensitivity to streamers and other off-limb structures, but has not been readily used by the forecasting community.

With a multi-channel system, different complementary pass-bands should be included, such as one to image hot plasmas (e.g. \SI{13.1} {\nano\meter}), one to image the quiet corona (\SI{19.5} {\nano\meter}), and one to image the chromosphere and cooler structures (\SI{30.4} {\nano\meter}). Although such pass-bands may be observationally complimentary, from a technical perspective it would be difficult to incorporate them in a single or two channel telescope. In general, different filters combined with mirror coatings, such as in a filter wheel design, are used to select different spectral pass-bands. The coatings and filters have to be complimentary. For example, pass-bands, such as those centred on \SI{17.2} {\nano\meter} and \SI{30.4} {\nano\meter}, can be included in a single coating design, as they fall in successive Bragg orders of the coating. This is not the case for other combinations, such as the \SI{13.1} {\nano\meter} and \SI{30.4} {\nano\meter} pass-bands. Different solutions could be envisaged: a pattern coating with a masked filter wheel, or superposed mono-band coatings. But, these solutions lead to reduced radiometric performances. Alternatively, a custom specific coating could be developed, but this produces development risks for the mission.

\begin{figure}
\centering
\subfigure{\includegraphics[width=0.48\columnwidth]{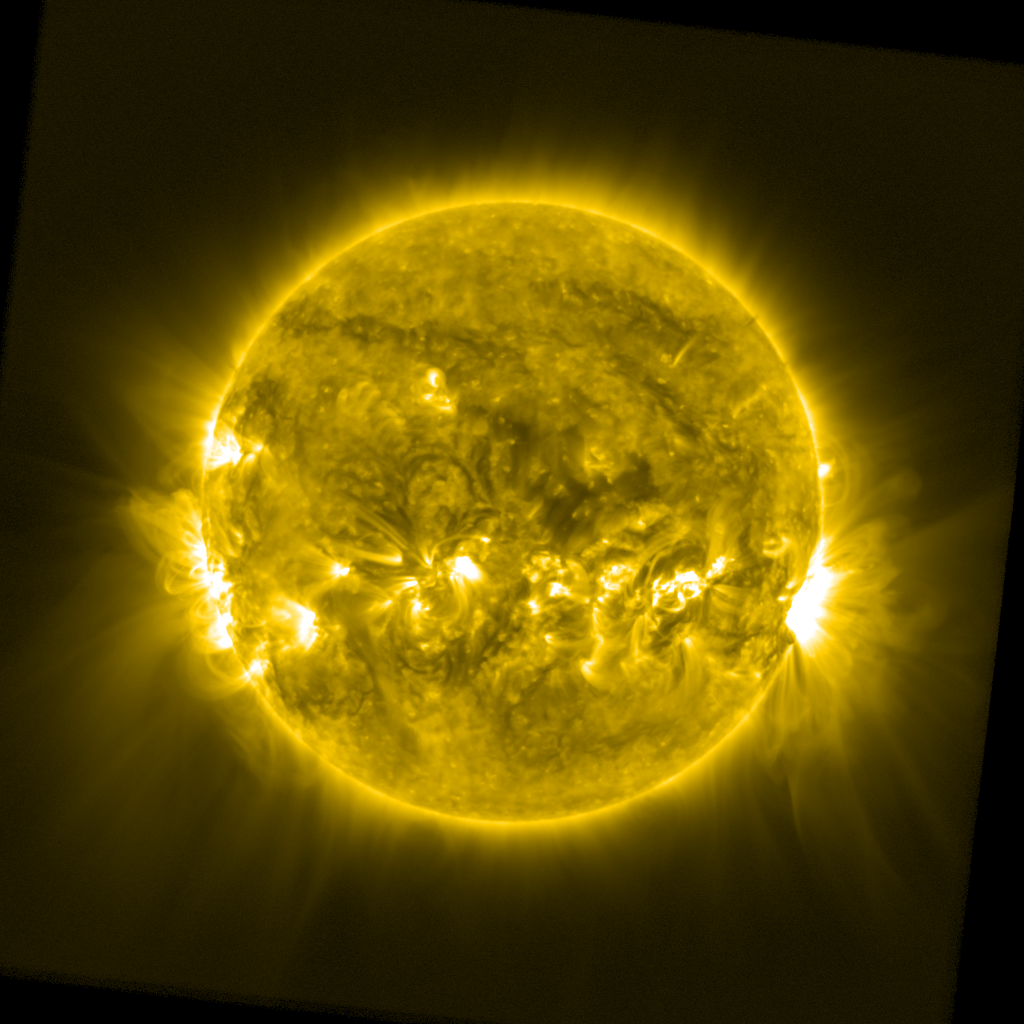}}\subfigure{\includegraphics[width=0.48\columnwidth]{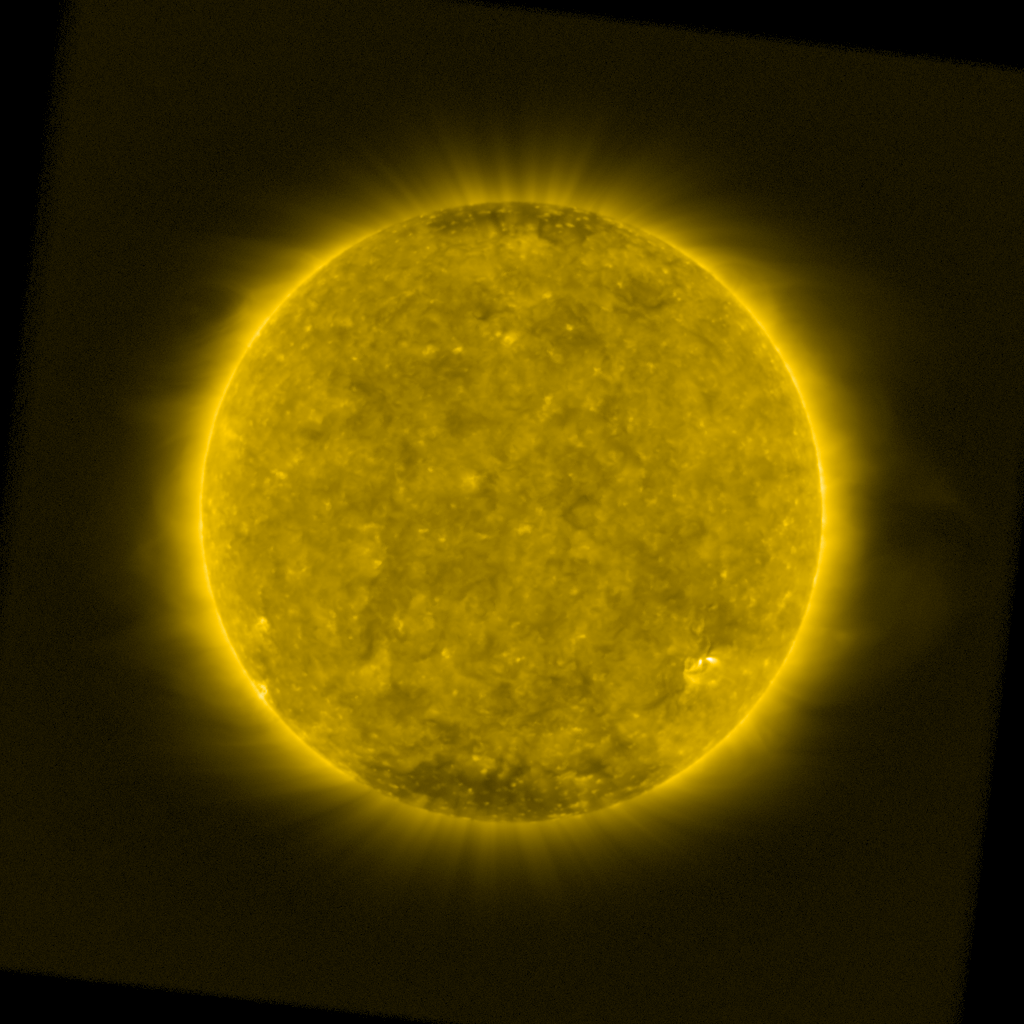}}
\caption{\small Two images of the Sun from the SWAP instrument on PROBA2, from 2014-Jan-01 at solar maximum (left) and from 2020-Jan-01 at solar minimum period (right). The images are constructed using a median stack of images over 100 minute blocks to suppress noise and enhance the off-limb solar signal.} 
\label{fig:largeFOV}
\end{figure}

The strong off-limb signal observed in the \SI{19.5} {\nano\meter} pass-band was another contributing factor for its selection, which is observed to diminish relatively slowly \citep{Tadikonda2019}. Off-limb EUV observations have become of increasing interest in recent years, where EUV imagers have been used to track structures in the off-limb lower and middle corona: the SWAP instrument on PROBA2 boasts one of the largest FOVs of any EUV imager currently monitoring the Sun, with a $54\times54$~arcmin FOV. Figure \ref{fig:largeFOV} shows two images of the solar atmosphere from the SWAP telescope on 2014-Jan-01 (solar maximum; left) and 2020-Jan-01 (solar minimum; right). The images are constructed using a median stack of images over 100 minute blocks. This way of constructing images suppresses random noise and one-time events such as energetic particle hits, yielding high signal-to-noise ratio images, which accentuate long lived structures in the far field of the images. These images have been used to reveal the large scale evolution of the EUV corona, including streamers and coronal fans \citep[see e.g.][]{Seaton2013b, Mierla2020}. The two images not only reveal the changing face of the Sun through the solar cycle, but also the extent of the solar atmosphere, seen off-limb. It's the complexity and structure of the extended corona which will influence solar activity such as eruptions, especially in their early phases.

Large FOV EUV images can also be used to show more dynamic structures such as the early phases of a CME. On 2017-Apr-03 the PROBA2 instrument was off-pointed by an additional 495 arcsec relative to the nominal Sun centred pointing, aligning it with the inner edge of the LASCO-C2 coronagraph. While off-pointed in this favourable position, the imager observed an eruption close to the limb, where it was possible to track the eruption through both the SWAP and LASCO FOVs, creating a set of observations spanning from the Sun out to 32 solar radii. The event is described in \citet{OHara2019}, where the eruption, dimmings, and EUV waves associated with the event were tracked. Other studies of eruptions using large FOV EUV observations include \citet{Sarkar2019} and \citet{Cecere2020}. The de-centred configuration of the LUCI FOV will allow it to produce continual observations of the extended EUV solar atmosphere, and provide important observations of the early development of eruptions, during the crucial acceleration phase.

\subsection{Material and Detector Decisions}
\label{sec:MaterialChoice}

Many of the material decisions related to the LUCI design were based on SWAP and EUI heritage, but one main difference comes in the material used for the optical bench and brackets. Two materials were considered, the first largely comprised of Invar, while the second was aluminium. Both have characteristics beneficial for supporting the LUCI mission. Invar offers a uniquely low coefficient of thermal expansion, whereas aluminium offers low electromagnetic interaction with the other sub-systems on the Lagrange satellite. Due to the relative stability of the thermal environment, the aluminium option was chosen, together with an active thermal control to compensate for possible expansion or contraction of the bench and brackets before and after launch, which could affect the optical layout.

The filters onboard LUCI, like SWAP and EUI need to be supported due to their size. Different approaches are available, which include: a mesh, a polyimide membrane, or a combination of both. Polyimide is strong but its EUV transmission is very low. The support mesh configuration offers a better transmission budget and a better mechanical resistance. However, the mesh can create a grid diffraction pattern: where the image of one bright point will produce a pattern of secondary points. Due to the compact design of LUCI, with its relatively small entrance pupil, a mesh filter support system was chosen with a \emph{thin} layer of polyimide, which has an achromatic transmission of approximately 82\%. With the LUCI design a shadowing effect will be created due to the focal plane filter and a diffraction pattern by the entrance filter. The pattern will be simulated and compensated for, as is performed with both EUI and SWAP observations.

Finally, the LUCI instrument will make use of an APS-CMOS detector in a backside thinned configuration. The main rationale behind the choice of detector is the heritage from the Solar-Orbiter EUI instrument \citep{Rochus2019}. CCDs have been used extensively in space-based observatories, offering a rich reliable heritage. When placed in a back-sided configuration, such as that used on AIA, they can even forgo a coating. However, the CMOS detector offers several advantages over a traditional CCD for an operational EUV imager, including: shutterless operation, where each pixel can be read out separately, greater radiation tolerance (no charge transfer), low power consumption, high operational temperature, high speed, high dynamic range, and non-destructive readout. The shutterless design is extremely advantageous for the operational mission design as it negates the need for a shutter mechanism, and a point of possible mechanical failure. The design also allows each pixel to acquire a high and low gain signal, allowing for the combination of both to increase the dynamic range. The lack of shutter will result in periodic off-point campaigns being required to perform calibration campaigns using the onboard calibrated diodes, without solar EUV signal and stray-light in the FOV.

\subsection{LUCI's Location}
\label{sec:LUCIlocation}

It's important to consider the benefits of sending an instrument all the way to the L5 point to monitor the Sun and space weather, both of which are normally monitored from the Earth perspective. The L1 point has obvious advantages for monitoring space weather. In-situ instruments experience the ambient conditions in the solar wind before they reach the Earth, and remote sensing instruments can see what is facing the Earth. However, the L1 position is situated a mere 1.5 million kilometres from Earth (1\% of the Sun-Earth distance), restricting the warning time from in-situ instruments, and the view of solar phenomena, that may soon affect the Earth, is restricted to a quarter of the Sun. HSSs and CIRs, linked to coronal holes seen in the lower solar atmosphere, sweep out following the Parker spiral. These structures co-rotate with the Sun, and would impact instruments at L5 several days before arriving at Earth \citep{Gopalswamy2011}. 

Another potential location considered for the Lagrange mission was the L4 Lagrangian point. From an EUV observing perspective, L4 is equally as good as L5 for monitoring structures travelling close to the Sun-Earth line, due to the optically thin nature of the observations. But, observations from the L4 perspective lack information on potential activity that might be rotating towards the Earth, such as activity related to coronal holes and active regions. The L5 position on the other hand sits ahead of the Earth from the perspective of solar rotation, and can observe the coming face of the Sun.

A disadvantage of observing purely from the L1 perspective is the difficulty in estimating the arrival time of eruptions at Earth. As eruptions move towards the Earth they're generally monitored in white light coronagraph observations, such as LASCO, where the Sun is occulted, and CMEs are observed as expanding halos \citep{Gopalswamy2010}. From this perspective it is not only difficult to ascertain the true speed, but at solar maximum, when multiple potential source regions may be present, the solar source can be difficult to identify in EUV observations. \citet{Rodriguez2020} used observations from the STEREO~B satellite when it was located near the L5 point to test our ability to forecast the arrival of a CME using observations from this position. They used observations of an Earth‐directed CME, seen as a partial halo by SoHO at L1. The STEREO~B observations allowed the CME to be tracked continuously, and by using the observations, in combination with methods to calculate the CME arrival time at the Earth (extrapolation, drag‐based model, and a magnetohydrodynamic model; see references within), they demonstrated that the estimate of the CME arrival time could be drastically improved by adding L5 data. \citet{Mierla2009} also used STEREO observations to track CMEs, but this time making stereoscopic reconstructions to infer the three-dimensional structure of the CMEs and their direction of propagation. EUV observations provided information about the source location, which was necessary to make accurate reconstructions. More recently, \citet{Mierla2013} used EUV observations from SWAP on PROBA2 and EUVI on STEREO to make three-dimensional triangulations of an erupting prominence. The triangulation technique was used to derive the true direction of the prominence propagation. With a satellite permanently positioned at the L5 point, in conjunction with observations from an L1 vantage point, more sophisticated reconstructions could be used as standard, to accurately track eruptions.

EUV and coronagraph observations made from the L5 perspective will not only help in characterising the kinematics and direction of eruptions, which are highly directional, but they will also help in identifying the source region. Measurements made of eruptions in large FOV EUV images will also help in characterising the acceleration phase of an eruption \citep{OHara2019, Zhang2006}. The advantages of identifying the origin of an individual eruption go beyond single events. As the mechanism triggering eruptions is not fully understood, the forecasting of flares and eruptions is often probabilistic \citep{Barnes2016}, meaning regions that have shown previous evidence of activity will have an increased probability of producing further activity. Thus, identifying the source region of flares and eruptions helps in identifying the sources of future activity. The L5 perspective will add an additional 60 degrees of observations to those acquired from L1, see Figure \ref{fig:LagrangePosition}, giving us the ability to monitor regions of the Sun before they're Earth effective.

\section{Conclusions}
\label{sec:Conclusion}

LUCI is a single channel imager with an EUV pass-band centred on \SI{19.5}{\nano\meter}, corresponding to emission produced mainly by highly ionised Fe~XII, with a peak temperature of 1.6~$\times$~10$^{6}$~K. LUCI will observe the Sun with a novel \emph{wide} 61.3~$\times$~42.7 arcmin FOV, through 2300~$\times$~1600 pixels, with a plate-scale of 1.6~arcsec per pixel.

LUCI is designed to be a compact (960~mm $\times$ 310~mm, $\times$ 195~mm), low mass ($\approx$ 15~kg), low power consumption ($\approx$ 10~W) solar monitor, to be located at the L5 Lagrangian point. To improve instrument robustness, and reduce the chance of operational and mechanical failure its heritage and design relies heavily on that of the SWAP and EUI instruments.

From the L5 perspective, with its novel off-pointed wide-FOV, LUCI will not only be able to monitor the sources of solar activity before they're Earth effective, by observing the solar disk before it rotates towards the Earth, but also monitor activity such as eruptions close to the Sun-Earth line, helping characterise them and their potential impact at Earth.

\begin{acknowledgements}
     The development of the LUCI instrument is currently funded as a consortium of the Belgian Federal Science Policy Office (BELSPO)  and the Swiss Space Office (SSO). The image in Figure~\ref{fig:SWAPoffpoint} was created using the ESA and NASA funded Helioviewer Project \citep[JHelioviewer,][]{Mueller2017}. SWAP is a project of the Centre Spatial de Liege and the Royal Observatory of Belgium funded by the Belgian Federal Science Policy Office (BELSPO), SWAP Carrington rotation data was obtained through the ESA science archive. AIA and HMI data was used courtesy of NASA/SDO and the AIA, and HMI science teams. LASCO data was used courtesy of SoHO/LASCO and the LASCO science team. SUVI data was used courtesy of GOES/SUVI and the SUVI science team. The authors would like to thank D. B. Seaton for help preparing the SUVI data used throughout, and B. Nicula for discussions relating to image recoding and C. Bizzi for proof reading. We also thank the anonymous referees for useful comments and suggestions.
\end{acknowledgements}

\bibliography{ms}

\end{document}